\documentclass[12pt,preprint]{aastex}
\def\msun{\thinspace M_\odot}

\slugcomment{}

\begin{document}

\title{Tidal Interactions of Red Giants with Environment Stars in Globular Clusters}
\author{Shimako Yamada\altaffilmark{1}, Atsuo T. Okazaki\altaffilmark{2} and  Masayuki Y. Fujimoto\altaffilmark{1}}
\altaffiltext{1}{Department of Physics, Hokkaido University, Kita-ku, Sapporo 060-0810, Japan} 
\altaffiltext{2}{Faculty of Engineering, Hokkai-Gakuen University, Toyohira-ku, Sapporo 062-8605, Japan} 


\begin{abstract}

We investigate the tidal interactions of a red giant with a main sequence in the dense stellar core of globular clusters by Smoothed Particle Hydrodynamics method. 
   Two models of $0.8 \msun$ red giant with the surface radii $20$ and $85 R_\sun$ are used with 0.6 or $0.8M_\sun$ main sequence star treated as a point mass.  
   We demonstrate that even for the wide encounters that two stars fly apart, the angular momentum of orbital motion can be deposited into the red giant envelope to such an extent as to trigger rotational mixing and to explain the fast rotation observed for the horizontal branch stars, and also that sufficient mass can be accreted on the main sequence stars to disguise their surface convective zone with the matter from the red giant envelope.  
   On the basis of the present results, we discuss the parameter dependence of these transfer characteristics with non-linear effects taken into account, and derive fitting formulae to give the amounts of energy and angular momentum deposited into the red giant and of mass accreted onto the perturber as functions of stellar parameters and the impact parameter of encounter.  
  These formulae are applicable to the encounters not only of the red giants but also of the main sequence stars, and useful in the study of the evolution of stellar systems with the star-star interactions taken into account.  

\end{abstract}


\section{INTRODUCTION}

In the core of globular clusters, it is thought that star-star interactions play an important role because of very high stellar density and of relatively low velocity \citep{hills76b}.  
   There is growing evidence for the modifications of stellar properties and populations under the influence of close encounters and collisions.  
  For example, the overabundance of low-mass X-ray binaries and millisecond pulsars is regarded as consequent upon tidal captures of an environment star by neutron stars, and upon the exchange encounters involving a neutron star \citep{fabian75,hills76a}:  
   the smaller relative frequency of red giants in the core is attributed to the deprival of their envelope during close encounters with environment stars \citep{djorgoski91, beer2004}, and blue-stragglers, which are main sequence stars more massive than the turn-off stars, may result from direct collisional coalescence and/or binary merge of two or more stars \citep[][see Mapelli et al. 2006, and Leigh et al. 2007 and references therein for recent works]{leonard89}; 
   see also reviews by \citet{bailyn95}, \citet{hut03} and \citet{ferraro06}.  
   In particular, the inflation of the number of known blue stragglers boosted by the observations using Hubble Space Telescope \citep{ferraro97, ferraro99, paltrinieri01, ferraro03, ferraro04, sabbi04, beccari06, warren06}, suggests that a significant fraction of stellar populations undergo such encounters with neighboring stars.  
   Recent observations with the {\sl Chandra} X-ray Observatory indicate a link between the numbers of X-rays binaries and the stellar encounter rates in globular clusters \citep{pooley03, pooley06}. 

Furthermore, there is a longstanding problem of large star-to-star variations in the surface abundances of light elements such as C, N, O, Ne, Mg and Al.  
   Some giants in globular clusters exhibit the anomalous surface abundances that cannot be explain in terms of the nucleosynthesis and material mixing in the stars within the current standard framework of stellar evolution \citep[e.g., see reviews by][]{kraft94, dacosta97}. 
   Since these anomalies are observed only in globular clusters but not from field giants in the Galactic halo, it is natural to search for their origin(s) in the differences between the environment in the globular clusters and in the Galactic halo, and hence, to consider them as a evidence of the star-star interactions. 
   In fact, \citet{fujimoto99} have proposed a scenario for the formation mechanism of these abundance anomalies involving hydrogen shell flashes in red giants, as a result of internal mixing, triggered by the deposition of angular momentum into their envelopes during a close encounter with other stars.  
   It is demonstrated that this extra-mixing model can reproduce the observed relationship such as correlation and scatter in the anomalous abundances of Na and Al and the Mg-Al anti-correlation \citep{aikawa01,aikawa04}.  
   
Recently, similar abundance variations are found among unevolved stars of turn-off and sub-giant \citep{gratton01}.  
   It is true that the variations and anti-correlations between CN and CH bands have been reported not only for giants but for stars of upper main sequence, which may be taken to suggest the presence of abundance anomalies in unevolved stars \citep{suntzeff91,briley92, cannon,cohen}.  
   It has been argued that these facts refute the evolutionary scenario that the abundance anomalies are produced during the evolution along the giant branch and favor the primordial scenario that the stars were born of gas already subject to the anomalous abundances \cite[e.g.,][]{sneden04}.  
   As a possible compromise, the recycling scenarios have been proposed that the inhomogeneity is due to the surface pollution by accreting the ejecta of anomalous abundances from the erstwhile AGB stars of intermediate masses \citep{thoul02}, or that the second-generation stars were born from the gas polluted with the ejecta of AGB stars \citep{dantona04,dantona04b}.   
   \citet{ventura01} argue that the burning at the bottom of convective zone in low metallicity AGB stars can produce the observed O-Na and Mg-Al anti-correlations. 
   On the other hand, \citet{Fenner04} cast doubt upon the relevance of AGB ejecta to the observed anomalies. 
   In addition, the scenarios involve serious difficulties both in the mass supply and in the overabundances of CN and s-process elements, attendant with the third dredge-up during the evolution;  
    the necessary amount of mass ejecta only to cover and disguise the surface convection of giants may well exceed the total envelope mass that can be ejected from the erstwhile AGB stars, and the enrichment of s-process elements is never observed \citep{James04}. 

For the evolutionary scenario, it can also be argued that the abundance anomalies are printed onto the surface of unevolved stars through the mass transfer during the close encounters with such giants that have already developed the abundance anomalies;  
   the anomalous abundances are themselves attributed to the deposition of angular momentum into the convective envelope of giants during prior close encounters with environment stars \citep{shimada03}.   
   The surface convection of population II main sequence stars contain mass of $3 \times 10^{-3} M_\odot$ near the turn-off \citep[e.g., see][]{fujimoto95}, and hence, the accretion of mass of this order may suffice to disguise their surface abundances with those of accreted matter.  
   The evolutionary scenario with star-star interactions during the red giant branch taken into account are free from the above difficulties and have a fair prospect of giving a satisfactory explanation to these inhomogeneous anomalies. 
   Recently, the helium production by this extra-mixing mechanism is discussed \citep{suda07} in relevance to the splitting of main sequence branch, observed from $\omega$ Cen \citep{Bedin04} and from NGC 2808 \citep{piotto07}. 

One of the aims of this paper is to investigate whether the orbital angular momentum can be transferred into the envelope of giant from the orbital motion and whether the main sequence star can accrete the envelope mass from giants enough to disguise their surface layer with the accreted matter through star-star interactions. 
   It is argued that the rotation-induced mixing requires (differential) rotation of $\sim 0.01$ times the local critical rate from the energetic viewpoint \citep[e.g., see][]{fujimoto99}, although the proper theory is yet to be established.  
   From the observations, the horizontal branch stars are known to display a bimodal distribution of rotation velocity with the fastest rotators at velocity $v_{\rm rot} \sin i \gtrsim 30 \hbox{ km s}^{-1}$ (where $i$ is the inclination angle of spin axis) on the cooler side of horizontal branch where $T_{\rm eff} < 15000^\circ$ K \citep{peterson83, peterson95, cohen97, behr00a, behr00b, recio02}.  
   If we neglect the angular momentum loss during the transition to the horizontal branch, such rapid rotations require the angular momentum corresponding to the rotation rate of an order of $\Omega \simeq 0.01 \Omega_{\rm RG}$ at the tip of red giants ($\Omega_{\rm RG}$ being the critical rotation rate at the surface).
  This poses a problem of the origin of angular momentum since in the low-mass stars, the angular momentum is effectively extracted by magnetic braking during the main sequence phase and by mass loss during the red giant phase \citep[e.g., see][]{recio02,suda06}. 

The star-star interactions have been proposed as the mechanism(s) to form the unusual stellar objects discussed above and studied by many authors. 
   Among the analytical approaches, \citet{fabian75} first presented an idea and evaluated the possibility that the low-mass X-ray binaries are produced through tidal dissipation during the two body encounters involving a neutron star or low-mass black hole. 
   \citet{press77} developed the linear perturbation theory of the two-body tidal capture mechanism to derive a general formula for the amount of orbital energy, deposited into the oscillatory modes of stellar envelope during a periastron passage.
   \citet{lee86} and \citet{mcmillan87} worked out the cross sections for the binary formation via tidal capture of a main sequence star and  \citet{mcmillan90} that of a red giant, respectively. 
   These studies are, however, limited to the linear regime and can not deal with the non-linear effects such as the mass transfer between the stars and the mass loss from the stars owing to large deformations of the stars by tidal force.  
   
In order to estimate the non-linear effects during the close encounters, numerical simulations are necessary.  
  Among the numerical approaches, most studies have been devoted to understanding the resultant offspring of the stellar interactions. 
   Simulations of tidal encounters have been performed for the various combinations of stars, 
   e.g., a main sequence and a red giant star \citep{benz91}:  a neutron star and a main sequence star or a red giant star in an attempt to explain the formation of the low-mass X-ray binaries and the millisecond pulsars \citep{davies92, rasio91, davies95, lee96}: main sequence stars in encounter and collision, aiming at the formation of blue stragglers \citep{lai93, lombardi02}:  
   and a red giant star and a neutron star in relation to the formation of pulsars or ultra-compact X-ray binaries \citep{rasio91, lombardi06}.
   These studies have been performed exclusively by using the smoothed particle hydrodynamics (SPH) method, except for the encounters involving a massive black hole and a star, which were calculated by using a three-dimensional Euler hydrodynamic code \citep{khokhlov93a, khokhlov93b}.
   Recently SPH simulations are applied to the evolution of a giant planet through the tidal interactions with a sun-like star \citep{faber04, ivanov04}. 

The former hydrodynamic simulations, especially those of the tidal encounters between a red giants and a main sequence star by \citet{davies91}, have been carried out with a relatively small number of SPH particles and for limited range of parameters.  
   Their results are thought to be subject to limitations arising from low-mass resolutions since mass involved in the interactions decreases as the periastron distance increases, and the SPH method may not give a valid description of such situations where mass scales are as small as that allotted to each particles.  
   In our problems of surface pollution, we deal with the accretion of mass $ \sim 10^{-3} M_\odot$. 
   Simulations with finer mass resolutions, and hence with larger particle numbers, are necessary to investigate such encounters involving the transfer of mass of this order.  
   It is also desirable to perform simulations for a wide range of parameters, such as the periastron distance, the red giant models in different evolutionary stages, and the mass of main sequence stars in order to obtain a realistic and general information about the characteristics of the tidal interactions.

In this paper, we first carry out simulations of tidal interactions between a red giant and a main sequence star by using SPH method. 
   We make a detailed analysis of the amounts of energy and angular momentum, transferred from the orbital motion to the oscillation and spin of red giant, and the amount of mass, lost from the red giant and accreted onto the main sequence stars.  
   Based on the numerical experiments, we then attempt to clarify the parameter dependence of these characteristics and to formulate the quantitative outcome as simple functions of the stellar parameters and impact parameter of encounters. 
   The present results are applied to investigate the relevance of the scenario that the star-star interactions give rise to the abundance anomalies observed among not only giants but also main sequence stars in globular clusters.  
   The derived formulae will be useful to perform simulations of dynamical evolution of stellar systems with the effects of stellar interactions taken into account.  

The organization of the paper is as follows. 
   In next section, we describe our numerical methods, including the set-up of the initial conditions, the models of the red giant, the treatment of accretion and the determination of the viscosity of red giant models. 
   In \S 3, we present the results from our simulations with the discussion of the non-linear effects of tidal interactions. 
 In \S 4, we derive fitting formulae for the energy and angular momentum deposited into the envelope of the red giants and the mass accreted onto the main sequence stars during the tidal encounters.  
  The conclusions follow in \S 5, with the discussion about the application to the globular clusters. 

\section{METHOD OF NUMERICAL COMPUTATIONS}

In the present work, we use the three dimensional, smoothed particle hydrodynamics (3D-SPH) code, originally developed by Benz \citep{benz90,benz90a}, and extended by \citet{bate95}.  
   The variable smoothing length is adopted with the hierarchical tree method, originally written by \citet{press86}, to make the list of particles in the closest neighborhood of the particles. 
   The kernel and the integrating method in our code are respectively the standard cubic-spline kernel and a second-order Runge-Kutta-Fehlberg integrator with individual time steps for each particle \citep{bate95}.

In our cases, the timescale of periastron passing is much shorter than the Helmholtz-Kelvin timescale in the envelope [$\tau_{\rm HK} \simeq 1.5 \times 10^4 (R_{\rm RG}/ 20 R_{\sun})^{-1} (L_{\rm RG}/ L_{\sun})^{-1}$ yr, where $R_{\rm RG}$ is the surface radius of red giant]. 
    Accordingly we assume the adiabatic relation for the gas in the envelope of red giants. 
    In actuality, the code takes into account the change of the entropy due to viscous dissipation, although it may have only a minor effect since we deal with the tidal interactions at large distance, not accompanied by large shock dissipation. 

Our SPH code uses the standard form of artificial viscosity with two free parameters $\alpha_{\rm SPH}$ and $\beta_{\rm SPH}$, which respectively control the strength of the shear and bulk viscosity components and that of a second-order, von Neumann-Richtmyer-type viscosity \citep{mona83}. 
   It is known that the linear artificial viscosity can be reduced to the Shakura-Sunyaev viscosity prescription in the continuous limit;
   \citet{Meglicki93} derived a relation which connects the viscous force with the linear artificial viscous parameter $\alpha_{\rm SPH}$. 
  If the density varies on a length-scale much larger than the velocity, the shear viscosity, $\nu$, is written in terms of the artificial viscosity parameter $\alpha_{\rm SPH}$, in the form:
\begin{equation}
\nu = (1/10) \alpha_{\rm SPH}c_s h,   
\end{equation}
   where $c_s$ is the isothermal sound velocity and $h$ is the smoothing length \citep{okazaki02}. 
   In the envelope of red giants, on the other hand, we may relate the shear viscosity to the eddy-viscosity of convective motions, $\nu_{\rm eddy}$, evaluated at 
\begin{equation}
\nu_{\rm eddy} = v_{\rm conv} \cdot l_{\rm mix}.   
\end{equation}
   where $v_{\rm conv}$ is the averaged velocity of convective elements and estimated from the mixing length theory with the mixing length, $l_{\rm mix}$. 
   For a red giant model of mass $0.8M_{\sun}$ and the metallicity $[{\rm Fe}/{\rm H}] = -1.5$, the eddy viscosity is found to be nearly constant around $\nu_{\rm eddy} \simeq 7 \times 10^{15} \hbox{ cm}^2 \hbox{ s}^{-1}$ in the envelope when the radius $\sim 20R_{\sun}$ and the luminosity $\sim 100L_{\sun}$ \citep[see e.g.,][]{suda06}. 
    Since $c_s \cdot h \simeq 7 \times 10^{17} \hbox{ cm}^2 \hbox{ s}^{-1}$ on average for the red giant models constructed with SPH code (see below), we may approximate the eddy viscosity with a choice of $\alpha_{\rm SPH} = 0.1$.   
    We perform the simulations with the two values of linear artificial viscosity parameter, $\alpha_{\rm SPH} = 1.0$ of common use and $\alpha_{\rm SPH} = 0.1$ in order to see the effects of viscous forces. 
    As for the non-linear artificial viscosity parameter, we follow the usual prescription and set $\beta_{\rm SPH} = 2 \cdot \alpha_{\rm SPH}$ \citep{bate95}.  

\subsection{Initial Conditions and Approximations}

Our simulations consist of two steps, i.e., we first make the initial models of red giants in hydrostatic equilibrium with SPH particles, and then, follow the encounter with a main-sequence star.  
   Our red giant models are constructed with a total of 50,000 SPH particles of equal mass in the envelope and the core approximated by an appropriate external potential, while the main-sequence star is treated as a point mass. 
   We start the encounter simulations by placing a red giant and a main sequence at a separation of $5 R_{\rm RG}$. 
   Their relative velocity at this distance is calculated from the relative velocity at infinity, assumed to be $v_{\infty} = 10 \hbox{ km s}^{-1}$ in this work, and the impact parameter. 
   The red giants are assumed to be not rotating initially.  
   We set the total mass at $M_{\rm RG} = 0.8 \msun$ and adopt two models at the different evolution stages, the one with the core mass $M_{\rm core}=0.32M_\odot$ and the surface radius $R_{\rm RG}=20R_\odot$ and the other with $M_{\rm core} = 0.48M_\sun$ and $R_{\rm RG}=85R_\sun$; 
   the mass of one SPH particle is $0.96 \times 10^{-5}$ and $0.64 \times 10^{-5} \msun$, respectively.  
   The former model is taken to have the same model parameters as by \citet{benz91} and \citet{davies91} who use 7,132 SPH particles of unequal masses, and the latter corresponds to the structure realized near the tip of the red giant branch.
   For the main sequence star, we take two different masses of $M_{\rm MS} = 0.6$ and $0.8 M_\sun$. 

\subsection{Red Giant Models}

Thee envelope structure of red giant can be reproduced by placing the envelope mass under the influence of the gravity of core, modeled as a sphere of uniform density, according to \citet{fujimoto92}. 
   By solving the equations of hydrostatic equilibrium with an additional gravity $g$ of the core, expressed as 
\begin{equation} 
g = \cases{ - GM_{\rm core} r / R_{\rm core}^3 & \qquad for $r \le R_{\rm core},$ \cr
- GM_{\rm core} / r^2  & \qquad for $r > R_{\rm core}$, \cr} 
\label{eq:red-gravity}
\end{equation}
    with the core radius $R_{\rm core}$, we determine the density distribution of red giant envelope; see appendix~A for detail.  
    For the equation of state, we assume the polytrope of $P = K{\rho}^{(1+1/N)}$ with the polytropic index $N = 1.5$, which corresponds to the adiabatic equation of state with the adiabatic exponent $\Gamma = 5/3$;
   the polytropic constant $K$ stands for the specific entropy of the monatomic ideal gas. 

\begin{figure}
\plotone{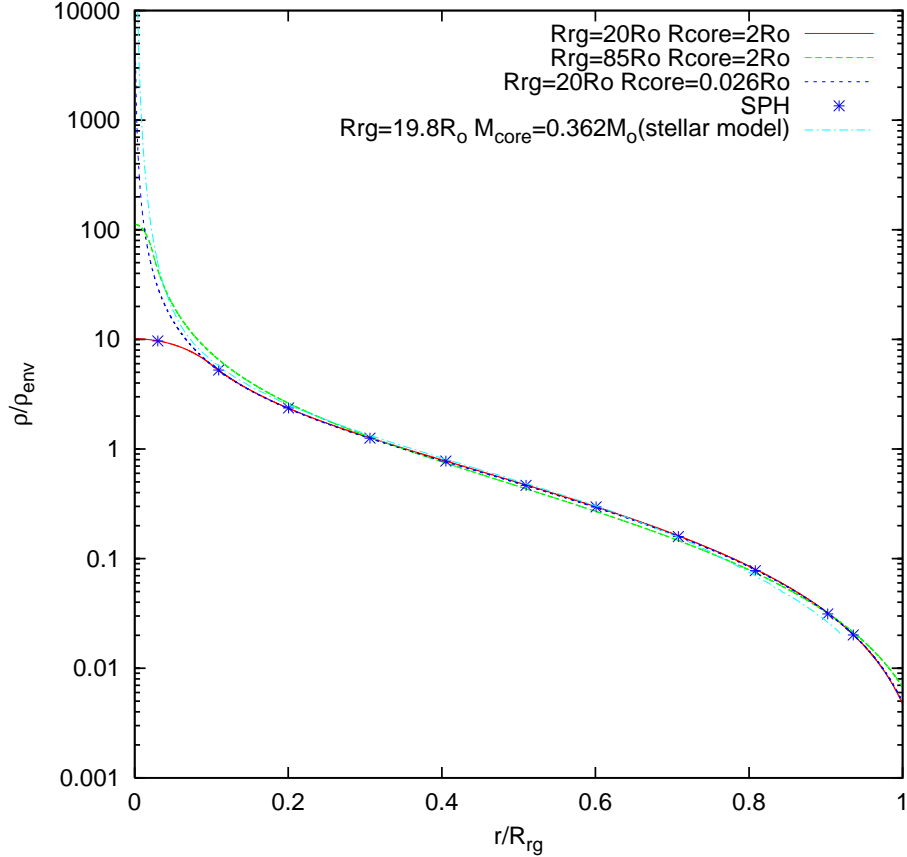}
\caption{
The distribution of density in the envelope of red giants, normalized with respect to the envelope mass and radius ($M_{\rm env}=R_{\rm RG}=1$) as a function of normalized radius.  
    Solid and long broken curves denote the models of the same core radius of  $R_{\rm core}=2R_{\sun}$ with the same total mass $0.8 M_\odot$, the different surface radius $R_{\rm RG}=20R_{\sun}$ and $85R_{\sun}$ and the different core mass $M_{\rm core} = 0.32 M_\odot$ and $0.48 M_\odot$, respectively, while broken curve denotes the model of a same surface radius $R_{\rm RG}=20R_{\sun}$ and a same core mass $M_{\rm core} = 0.32 M_\odot$ with the same total mass $0.8 M_\odot$ and the different core radius $R_{\rm core}= 0.026R_{\sun}$, for comparison. 
   Dash-dotted curve denotes a stellar model of $R_{\rm RG}=19.8R_{\sun}$ and $M_{\rm core} = 0.362 M_\odot$, taken from the evolutionary calculation of a star with mass $0.8 \msun$ and the metallicity $[{\rm Fe}/{\rm H}] = -1.5$ by \citet{suda06}. 
   Asterisks represent the density distribution of the SPH model with $R_{\rm RG}=20R_{\sun}$ for the radii at intervals of $2 R_\odot$ with two additional ones near the center and surface. 
\label{fig:redgiant}}
\end{figure}

Figure~\ref{fig:redgiant} shows the density distributions in the envelope of red giants, thus obtained, for the models with different surface radii of $R_{\rm RG} = 20 R_\odot$ and $R_{\rm RG} = 85 R_\odot$. 
    When the radius and the density are normalized with the surface radii $R_{\rm RG}$ and the envelope density $\rho_{\rm env} = M_{\rm env} / R_{\rm RG}^3$, two density distributions become nearly identical except inside of the core, which is a feature of red giant structure unless the mass in the envelope is much smaller than the core mass \citep{fujimoto92}.  
   In this figure, we also plot the density distribution in the red giant by taking the model from the evolutionary calculation \citep{suda06} for comparison, which exemplifies that the analytic models can reproduce the envelope structure of red giants very well. 
   Moreover, we compare the models with different core radii of $R_{\rm core} = 0.026 R_\sun$ and $2 R_\sun$ to demonstrate that the assumed core radii hardly affect the structure outside the core of $r > 2 R_\sun$, and in particular, in the outer envelope that may take a main part in tidal deformations, while the central density differs greatly by a factor of $4.6 \times 10^5$.  
    In our simulations, we therefore set $R_{\rm core} = 2 R_\odot$ to reduce the amount of particles injected into the innermost region. 
   The initial models of red giants for the SPH simulation are constructed by distributing particles according to these envelope solutions, and then, by relaxing them into hydrostatic equilibrium with an artificial damping force on the particles.
   The relaxed distribution of SPH particle is also shown in the figure;
   it reproduces the structure of red-giant envelope very well except for the very surface layer of mass less than $\sim 0.0001 \msun$ because of mass resolutions, where the variable smoothing length, h, $\sim0.1R_{RG}$. 

\subsection{Accretion onto the Main Sequence Star} 

We assume that the main-sequence star, treated as a point mass perturber, accretes any SPH gas particles that enter within the accretion radius, $r_{\rm acc}$, which is defined as half the Roche-lobe radius, $R_{\rm L}$, calculated under the assumption of a circular orbit at a instantaneous separation, $D$, between the main sequence and the red giant, and given by; 
\begin{equation}
r_{\rm acc} = 0.5 R_{\rm L} = 0.5 D ( 0.38 + 0.2\log q )
\label{eq:Roche-lobe}
\end{equation} 
   in which the expression for $R_L$ is valid for the mass ratio $ 0.3 < q = M_{\rm MS} / M_{\rm RG} <20 $ \citep{Paczynski71}. 
   A factor of 0.5 is adopted in order for the main sequence star not to artificially accrete unbound particles that happen to enter its Roche lobe. 
   We have confirmed that the number of accreted particles is nearly the same with a smaller accretion radius $r_{\rm acc} = 0.1 R_{\rm L}$.  
   As for the accreted SPH particles, the mass, momentum and angular momentum that they carry are added to the point mass of main sequence stars. 

\section{RESULTS OF SIMULATIONS }

We have carried out 25 simulations of tidal encounter with a red giant of mass $0.8 M_\odot$ and a main sequence star by varying the impact parameter, $b$, for eight sets of parameter combinations with the two different red giant models, the two different main sequence star and the two choices of artificial viscosity parameters.   
   The model parameters are summarized with model identifiers in Table~\ref{tab:model=alpha}. 
   We adopt relatively heavy main-sequence stars of mass, 0.8 and $0.6 \msun$, based on the fact that the mass segregation may proceed to enhance the abundance of relatively massive stars in the cluster cores where the close encounters are expected to occur more frequently because of larger stellar density.  
  In this table we give the periastron distance, $r_p$, instead of the impact parameter $b$, which is given for a hyperbolic orbit as 
\begin{equation}
b^2 = r_p^2 [ 1 + 2 G (M_{\rm RG} + M_{\rm MS})/r_p v_{\infty}^2],     
\end{equation} 
   where $v_{\infty}$ is the relative velocity before the encounter and set to be $v_{\infty} = 10 \hbox{ km s}^{-1}$ in the present work. 
   We also define the ratio, $\eta$, between the critical angular velocity, $\Omega_{\rm RG}$, of rotation at the initial surface of red giant and the angular velocity, $\Omega_{\rm pass}$, for the circular orbit at the periastron distance, $r_p$, as a measure of the closeness of encounter; 
\begin{equation}
\eta = \frac{\Omega_{\rm RG}}{\Omega_{\rm pass}} = \left(\frac{M_{\rm RG}}{M_{\rm RG}+ M_{\rm MS}}\right)^{{1}/{2}} \left(\frac{r_p}{R_{\rm RG}}\right)^{{3}/{2}}.  
\label{eq:eta}
\end{equation}
   following \citet{press77}.  
   In addition, we give the characteristic results of simulations, the energy and angular momentum transferred into the red giants from the orbital motions, and the masses, accreted onto the main sequence stars and lost from the systems; 
   also listed are the periods, semi-major axes and eccentricities of orbital motions for the models that yield bound systems and the models that end up with positive orbital energy are denoted as fly-by.  

\begin{figure}
\plotone{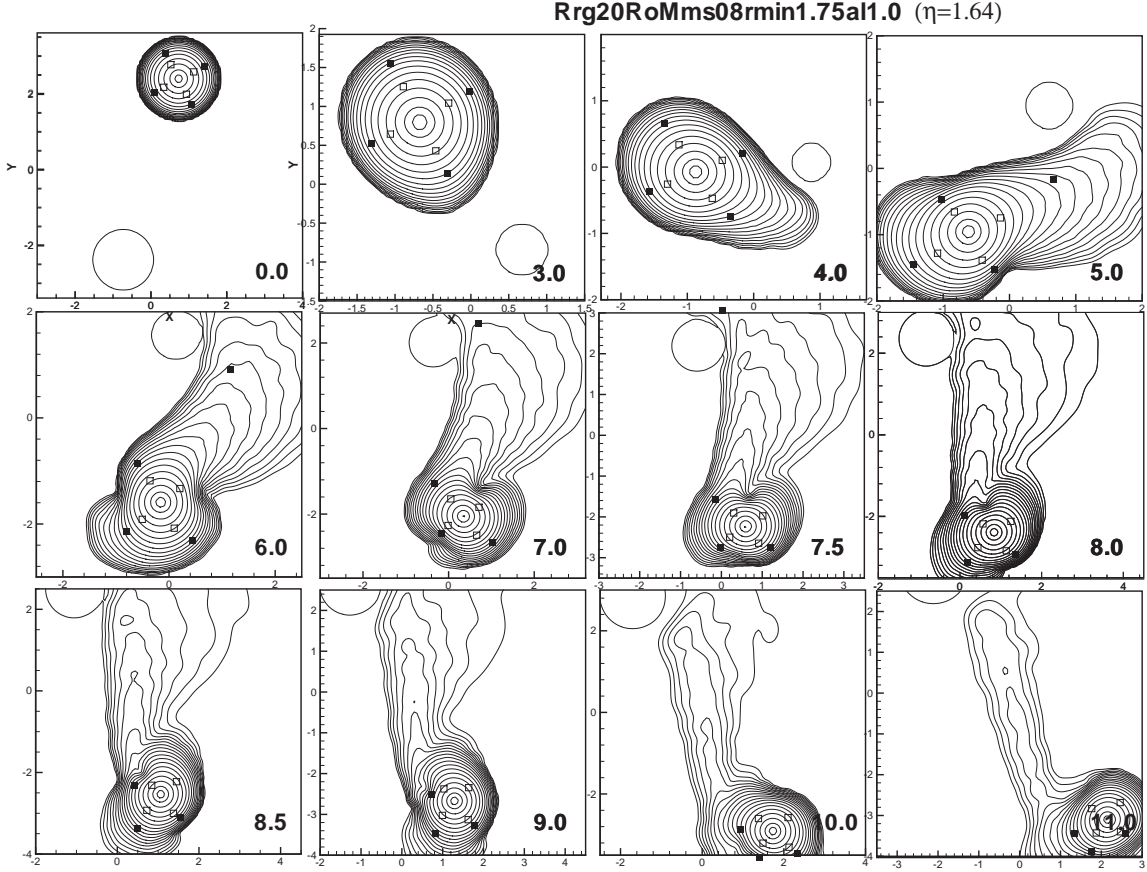} 
\caption{
Variations in the surface density, projected on the orbital plane, during the encounter for case a8rg1 ($R_{\rm RG}=20R_{\sun}$, $\eta=1.64$, $M_{\rm MS}=0.8M_{\sun}$, $\alpha=1.0$).  
   Each panel shows a snapshot of contour lines at intervals of 0.2 dex in the logarithmic scale over the range of 10-0.001 times the average surface density, $\Sigma_{\rm env} = M_{\rm env}/ \pi R_{\rm RG}^2$.  
   The origin of coordinate is set at the center of mass and numerals in the right bottom corner indicate the time in units of dynamical timescale, $\tau_{\rm RG} = \sqrt{ {R_{\rm RG}^3} / {GM_{\rm env}}}$.     
  Filled and open squares indicate representative SPH particles initially located on the shells that contain 95\% and 70\% of total mass including the core mass in the interior, respectively.
\label{fig:snapshota8rg1}}
\end{figure}

Figures~\ref{fig:snapshota8rg1} gives the snapshots showing the variations of surface density, $\Sigma$, projected on the orbital plain for Model a8rg1 ($R_{RG}=20R_\sun$, $M_{MS}=0.8M_\sun$, $\alpha_{SPH}=1.0$, $r_p/R_{RG}=1.75$ or $\eta=1.64$); 
   the contours, separated by 0.2 dex, are plotted in the range of $10 - 0.001$ times the average surface density, $\Sigma_{\rm env} = M_{\rm env}/ \pi R_{\rm RG}^2$, and open circle denotes the accretion radius of the main sequence star. 
   Numerals in right-bottom corner give the elapsed time from the onset of simulation in units of dynamical timescale, $\tau_{\rm RG} = ({R_{\rm RG}^3/G M_{\rm env}})^{1/2}$, defined with the envelope mass of red giant as in \citet{davies91}.  
   On each panel open and filled squares mark the gas particles, initially situated on the two separate shells on the orbital plane, as the indicators of stellar rotation. 

As the main sequence star approaches, the tidal bulge is raised on the surface layer of red giant and grows toward the main sequence star.  
   The oscillations of $l=2$ f-modes are predominantly excited as predicted from the linear perturbation theory.  
   In the outer shells of a few 10 \% in mass fraction, the deformations greatly elongated toward the perturber develop into non-linear regime, as seen from filled squares;  
   in the interior, on the other hand, the perturbations remain small in linear regime, as observed from the location of open squares, and the gas almost stay at rest in the still deep interior.
   The outer deformations cannot keep pace with the motion of perturber as it is accelerated because of the timescale of passage of perturber comparable to that of oscillatory motions and of initially small rotation rate of the red giant.
   The lag of tide develops as the perturber approaches to the periastron.  
   Later around time of $6.5 \ \tau_{\rm RG}$ after the periastron passage, the gas streaming out of the red giant starts to accrete onto the main sequence star, now separated by $D \simeq 4 R_{\rm RG}$. 

\begin{figure}
\plotone{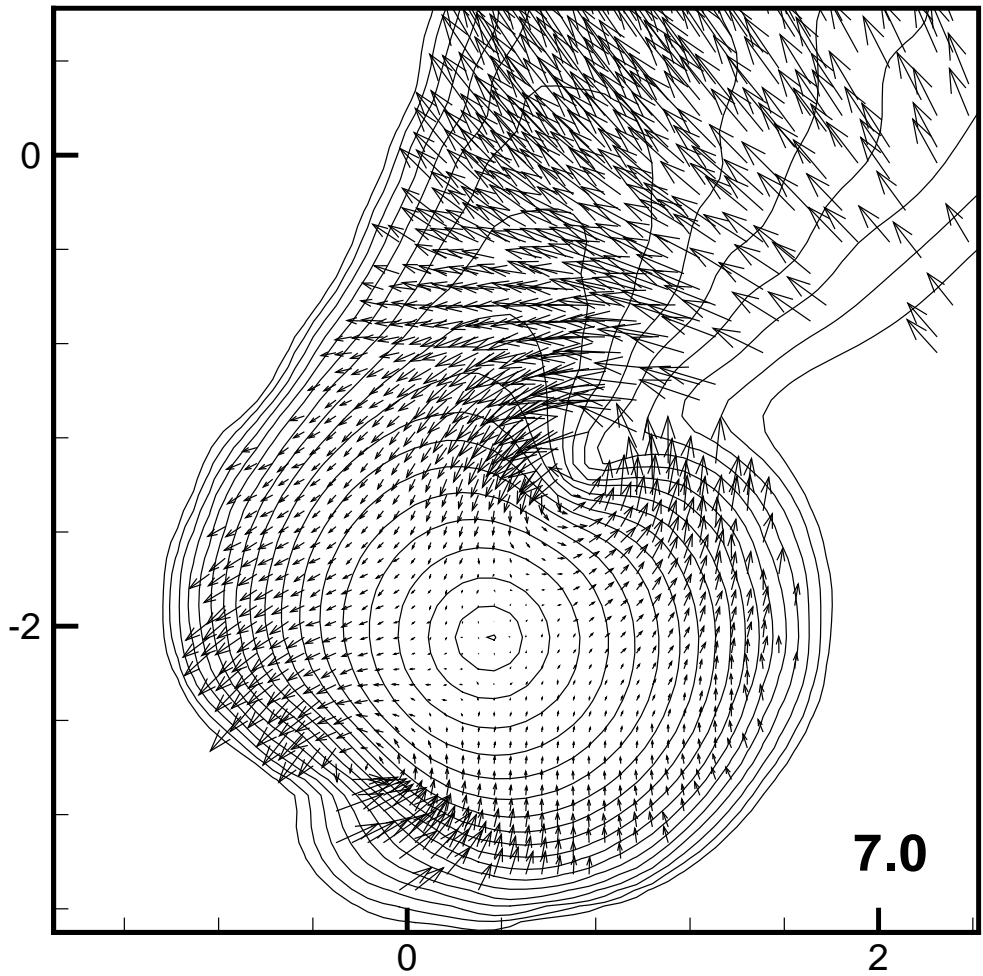}
\caption{
The velocity vectors, plotted on the density contour map of panel at time $7.0 \ \tau_{\rm RG}$ in Fig.~\ref{fig:snapshota8rg1}.
\label{fig:eddy-structure}}
\end{figure}

At the same time, there appears an interesting non-linear feature in the vicinity of the surface of the red giant star.
   As a result of the $l=2$, $f$-mode oscillations, a density ditch is formed near the interface of upward and downward motions, as seen from the panels in the middle row.
   Figure~\ref{fig:eddy-structure} shows an enlarged picture of the velocity structure when the ditch is formed.    
   Since the expanded mass elements gain larger tidal torque than the compressed mass elements, the former overtakes the latter while contracting to form an eddy-like velocity structure of counter-clockwise rotation.   
   The rotation of Lagrangian shells and the resultant deposition of angular momentum into the red giant star due to the tidal torque proceed spectacularly in the surface region, dominated by the non-linear effects.    
   Such features can no longer be the case in the linear theory, although the mass involved in the non-linear deformations is small, as seen from the movement of open squares.   
   The similar non-linear effects are reported by \citet{khokhlov93a}, who study tidal encounters between a polytropic star and a black hole, although they assume large periastron distance, and hence, the deformations remain nearly axial-symmetric without mass transfer, which is different from our case. 

This model illustrates an example that ends with the formation of a bound system after the encounter, as seen from Table~\ref{tab:model=alpha}. 
   In Model e8rg1 with the same parameter but for the smaller shear viscosity of $\alpha_{\rm SPH} = 0.1$, the deformations are identical during the earlier phase of time $0 \sim 5 \tau_{\rm RG}$ with those in Fig.~\ref{fig:snapshota8rg1}, and the effect of smaller viscosity is plainly discernible only in deeper ditch that develops after 6 time units.   
   Accordingly, there are only small differences in the results in Table~\ref{tab:model=alpha}.  
   This suggests that the transfer of energy and angular momentum is attributed solely to the phase-lag of the deformations, dynamically generated in the red giant envelope behind the perturber passage.   
   The interactions to exchange these quantities predominantly occur near the periastron passage of the smallest separation while the deformations are still growing.  
   On the other hand, the effects of viscosity become important only after the deformations contract to generate a strong shear near the stellar surface, and hence, the value of viscosity hardly affects the transfer characteristics. 

The models of larger impact parameters such as Model c8rg1 of $\eta = 2.4$ result in a fly-by encounter. 
   The non-linear deformations in the outer shells of red giant are weaker than those of the model of closer encounter $\eta=1.64$ in Fig.~\ref{fig:snapshota8rg1}.  
  Because of slower angular velocity of perturber, the tidal bulge stretched out toward the perturber is relatively slimmer to form a chimney like structure. 
   The ditch and the eddy-like structure, generated on the surface, are also in smaller scales. 

The encounter with a less massive perturber results in smaller transfer characteristics, a part of which is due to a larger periastron passing time, $\eta$, when compared at the same periastron distance.  
   Model c6rg1 of $M_{\rm MS}=0.6M_{\sun}$ and $r_p =2.00 R_{\rm RG}$ ($\eta = 2.14$) exhibits a similar chimney like structure as Model b8rg1 of the massive perturber with the same periastron distance, but the surface deformations are slightly smaller because of larger $\eta$.  
  The overall transfer characteristics lie between those of two massive perturber models, Model b8rg1 with $\eta = 2.00$ and Model c8rg1 with $\eta = 2.40$.  

The encounters with red giants at later evolutionary stages are exemplified by the models of the larger radius, $R_{\rm RG}=85R_\sun$.  
   When models with the same encounter closeness parameter $\eta$ are compared, the overall characteristics of interaction, i.e., the structure of mode oscillations, the development of deformations and the accreted process, are very similar in all models, despite of the large difference in the physical distance scales. 
   This is attributable to the nature of envelope structure of red giants, i.e., to the self-similar density structure, as shown in Fig.~\ref{fig:redgiant}.  
   The resultant transfer characteristics also turn out to be very similar when we subtract the effects of smaller envelope mass.  

The overall features of gas streaming are as follows. 
   For small periastron distances as in these encounters, the non-linear effects in the tidal interactions are important and the surface density profile becomes highly asymmetric in the outer shells of red giants.  
   As the perturber approaches to periastron, the tidal bulge is exited in the red giant and elongated, first directed to the perturber.  
   Since it cannot catch up with the motion of the perturber because of initially slow rotation, gas particles from the surface of red giant chases after the perturber to gain the energy and angular momentum. 
Some of them eventually get captured by the gravitational potential of the perturber after periastron passage.  
   On the other hand, most of gas involved in the tidal bulge falls back onto the red giant with gained angular momentum, which produces a non-linear feature of eddy-like structure in the surface region.  
   Even in the case of binary formations, the orbit is highly eccentric, and hence, as two stars go away from each other the tidal bulge becomes slender.  

\subsection{Non-Linear Deformations and The Evolution of Differential Rotation}

The transfer of orbital energy and angular momentum to the red giant is characterized by the following three time-scales;  
  (i) the periastron passing time scale of main sequence star, $\tau_{\rm pass}(= 1/ \Omega_{\rm pass})$, which is related to the variation of the external perturbing force: 
  (ii) the dynamical time-scale of the envelope of red giant, $\tau_{\rm dyn} (=1/\Omega_{\rm RG})$, which is related to the stellar oscillations in response to the external force: 
  (iii) the viscous time-scale in the envelope of red giant, $\tau_{\rm vis} (=R_{\rm RG}^2 / \nu_{\rm eddy})$, which is caused by the convective eddy in the red giant envelope.  
   In our case,  $\tau_{\rm vis} \ll \tau_{\rm pass} \lesssim \tau_{\rm dyn}$.  
   For a slowly rotating red giant, the tidal bulge tends to fall behind the accelerated motion of perturber, and hence, a tidal lag is formed dynamically to carry the energy and angular momentum from the orbital motion into the oscillatory motions and rotation of red giant envelope.  
   The viscosity plays secondary roles in the transfer of energy and angular momentum, as seen in Table~\ref{tab:model=alpha}. 
   The accreted mass depends little on the assumption of viscosity either.  
   Furthermore, we see that the accreted mass takes nearly the same values regardless of the difference in the red giant models when compared among the models of the similar values of $\Delta {E} / ( G M_{\rm RG}^2 / R_{\rm RG}$), instead of the encounter closeness parameter $\eta$.   
   This is indicative that the mass accretion is determined by the same process of energy deposition process.  

\begin{figure}
\plotone{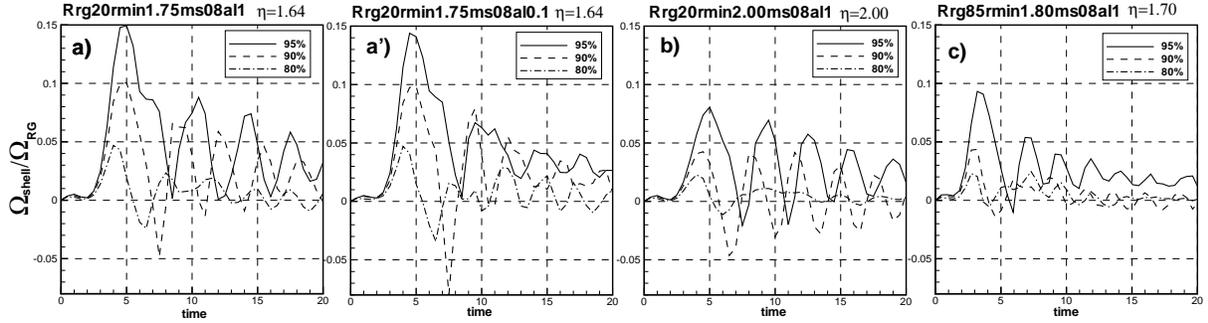}
\caption{
Time variations of angular velocity, $\Omega_{\rm shell}$, averaged over particles in Lagrangian rings on $z = 0$ orbital plain, initially located at the shells, the interior of which contains 95\%, 90\% and 80\% mass of red giant (including the core mass);  
   the vertical axis is the angular velocity normalized with respect to the Keplerian value at the initial surface ($\Omega_{\rm RG}= \sqrt{G M_{\rm RG} / R_{\rm RG}^3}$) and the horizontal axis is time in units of $\tau_{\rm RG}= 1 / \Omega_{\rm RG}$). 
   Leftmost panel (a) for Model a8rg1: the second and the third panels (a$^{\prime}$ and b) for Models e8rg1 and b8rg1, which differ from the first one in the viscous parameter and in the encounter closeness parameter, respectively; rightmost panel (c) for Model b8rg2 with the red giant model of larger surface radius. 
\label{fig:omegashell-time}}
\end{figure}

Figure~\ref{fig:omegashell-time} shows the time variations of angular velocity, $\Omega_{\rm shell}$, relative to its Keplerian angular velocity, $\Omega_{\rm RG} = (GM_{\rm RG}/R_{\rm RG}^3)^{1/2}$, averaged over the gas particles in the three Lagrangian rings on the orbital plane, initially located at the shells which contain the mass of red giant (including the core mass) by 95\%, 90\%, and 80\%, respectively.
   The tidal torque excites the oscillations of $\Omega_{\rm shell}$ of periods $\sim 2 \tau_{\rm RG}$, corresponding to the $l=2$, $m = \pm 2, f$-mode.  
  As the perturber approaches, the oscillations develop precipitously to reach the maximum strength near the periastron passage (at $t \simeq 4.5 \tau_{\rm RG}$ and $3.0 \tau_{\rm RG}$ for the models with $R_{\rm RG} = 20 R_\odot$ and $85R_\odot$, respectively) with greater amplitudes and longer durations in outer shells.
   Accordingly, the turnover of rotation velocity delays in the outer shell, and at the same time, the mean rotation velocity of oscillation increases in prograde direction, reflecting the injection of angular momentum due to the tidal torque, which is stronger in outer shells.  
   The amplitudes and mean-values of oscillations are greater for closer encounter.   
   Between the different red giant models with the same closeness parameter $\eta$, the model with the larger radius entails smaller variations in the rotation rate, normalized with respect to the characteristic rotation rate, $\Omega_{\rm shell}/\Omega_{\rm RG}$;  
   this is true even if we take into account the fact that the shell of the same mass fraction is 50\% deeper in the envelope because of smaller envelope mass.  
   It should be noted however that the net amount of angular momentum transferred is larger, though slightly, for the red giant model of larger radius because of larger critical angular momentum ($R_{\rm RG}^2 \Omega_{\rm RG})$. 

As for the effects of viscosity on the time variations of $\Omega_{\rm shell}$, the largest one appears in the difference in the minimum value after the periastron passage;  
   for the smaller viscous parameter, it decreases to be smaller, and along with the phase delay of outer shells, the eddy-like structure of counter-clockwise flow becomes stronger, as stated above. 
   The time of minimum $\Omega_{\rm shell}$ coincides nearly with the time of the strongest eddy in Figs.~\ref{fig:snapshota8rg1}.  
   Stronger shear produced by the eddy-like structure in turn enhances the dissipation and inward transport of angular momentum; 
   in panels a and a$^{\prime}$, we see the amplitudes of oscillations in the outer two shells get smaller in the second and later cycles for the models of smaller viscosity, though the overall similarity holds , in particular, in the shifts of mean-values of oscillations. 

\begin{figure}
\plotone{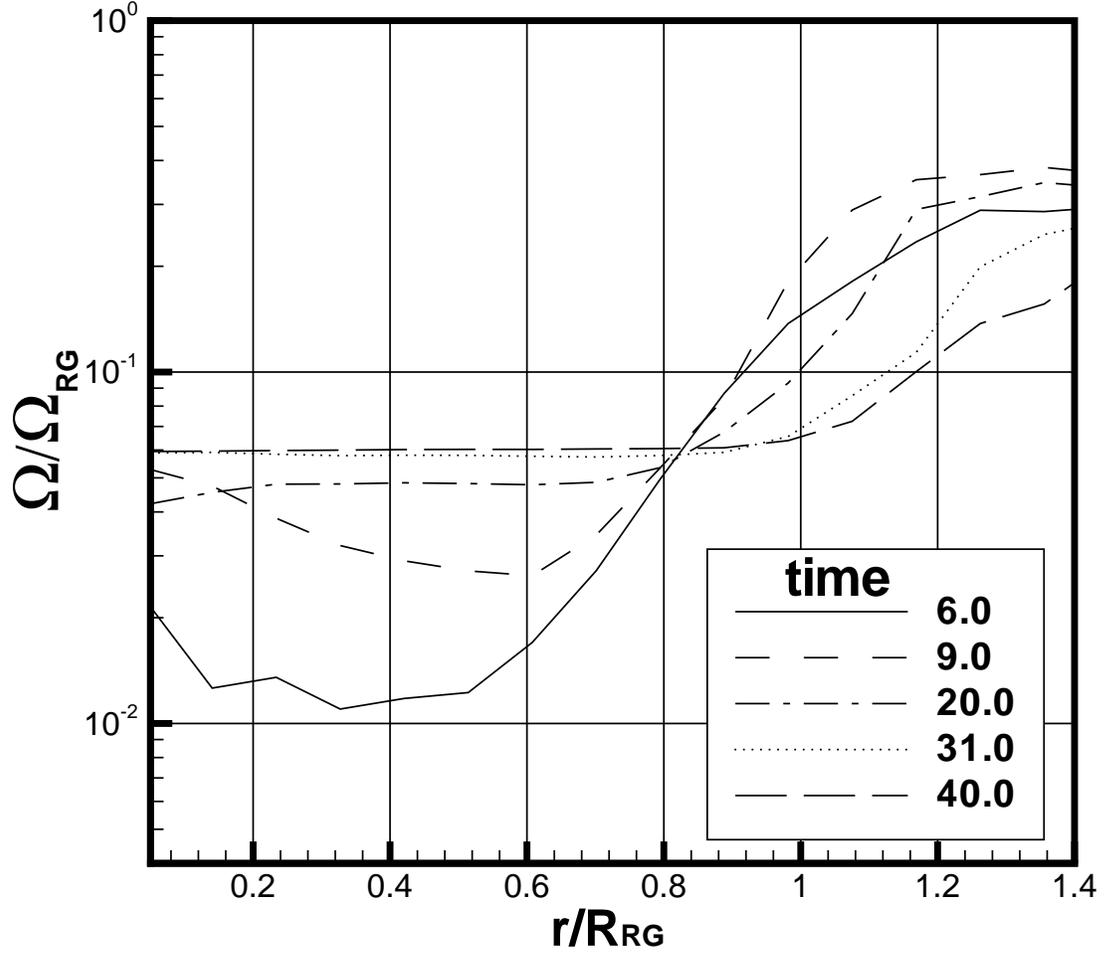}
\caption{The evolution of radial distribution of angular velocity, averaged over the particles that reside temporally between the cylinders, separated by $0.1R_{\rm RG}$ on the orbital plane, for model b8rg1; 
   $\Omega_{\rm RG} (= \sqrt{G M_{\rm RG}/R_{\rm RG}^3}$) is the critical rotation velocity at the initial surface of red giant, and the time is designated in the box in units of dynamical time of $\tau_{\rm RG} = \sqrt{R_{\rm RG}^3/GM_{\rm env}}$.
\label{fig:omega-radius}}
\end{figure}

Figure~\ref{fig:omega-radius} illustrates the evolution of the radial distribution of angular velocity, averaged over the gas particles between cylinders, perpendicular to the orbital plane, with the outer and inner radii, separated by $0.1R_{\rm RG}$.  
   The outermost layer is first accelerated and pulled most outwards to run after the perturber of the angular velocity, $\Omega_{\rm pass}$, at periastron passage ($\simeq 0.61 \Omega_{\rm RG}$ around the time $\sim 6 \tau_{\rm RG}$). 
   Then the deposited angular momentum is redistributed gradually into the interior. 
   By $\sim 40 \tau_{\rm RG}$, the most of interior up to the radius $R \simeq 1.1 R_{\rm RG}$ tends to rotate uniformly while there remains differential rotation in the outer expanded layer of lower density.  
   For the smaller viscosity parameter of $\alpha = 0.1$, the viscous process works slightly more slowly and a uniform rotation is reached only in the interior of $R \lesssim 0.8 R_{\rm RG}$ by $\sim 40 \tau_{\rm RG}$ with stronger differential rotation in the outer shells.  
   The timescale for transfer of angular momentum inside the red giants is shorter than that estimated from the eddy viscosity, the latter of which is $\sim$ 1 yr and 10 yr for the case of $\alpha_{SPH}=1.0$ and $\alpha_{SPH}=0.1$ respectively, while $\tau_{RG} = 0.02$ yr.  
   This also indicates that the non-linear effects, seen from Fig~\ref{fig:eddy-structure}, must mainly contribute to the redistribution of the angular momentum. 
    
\subsection{Transfer of Energy and Angular Momentum and Mass Accretion }

We evaluate the change in the orbital energy, $\Delta E$, from the difference of the kinetic and potential energy as  
\begin{eqnarray}
\Delta E (t) = \frac{1}{2} \mu_0 v_{\infty}^2 - \left( \frac{1}{2}\mu v^2 - \frac{GM_{\rm RG} M_{\rm MS}}{r} \right),
\end{eqnarray}
   where $\mu$ is the reduced mass: $r$ and $v$ are the relative distance and velocity, respectively, and the subscript 0 denotes the quantities before the encounter. 
   We may well assume that the change in the orbital energy is equal to the energy transferred to the red giant since the energy carried away by the particles that escape from the system is much smaller and escaped mass is smaller than accreted mass by nearly an order of magnitude as seen from Table~\ref{tab:model=alpha}. 

The total angular momentum, $\Delta L$, transferred into the red giant is estimated by summing up the specific angular momentum for all the particles constituting the red giant envelope around the core;
\begin{equation}
\Delta L (t) = \sum_i m_i ({\textit {\textbf r}}_i - {\textit {\textbf r}_{\rm RG}})\times({\textit {\textbf v}}_i - {\textit {\textbf v}_{\rm RG}}),
\end{equation}
   where $m_i$, {\textit{\textbf r}}${}_i$ and {\textit{\textbf v}}${}_i$ are respectively the mass of the i-th gas particle and its position and velocity vectors, and {\textit{\textbf r}}${}_{\rm RG}$ and {\textit{\textbf v}}${}_{\rm RG}$ are the position and velocity vectors of the core of red giant, respectively. 
   We exclude from the summation the gas particles which have accreted onto the main sequence and those which have escaped from the system;    
   the latter particles are defined as satisfying the following two conditions;
   (1) the total energy, i.e., the sum of the thermal, kinetic and potential energy, of a gas particle is positive and 
   (2) the radial velocity is positive when measured from the center of mass.  

\begin{figure}
\plotone{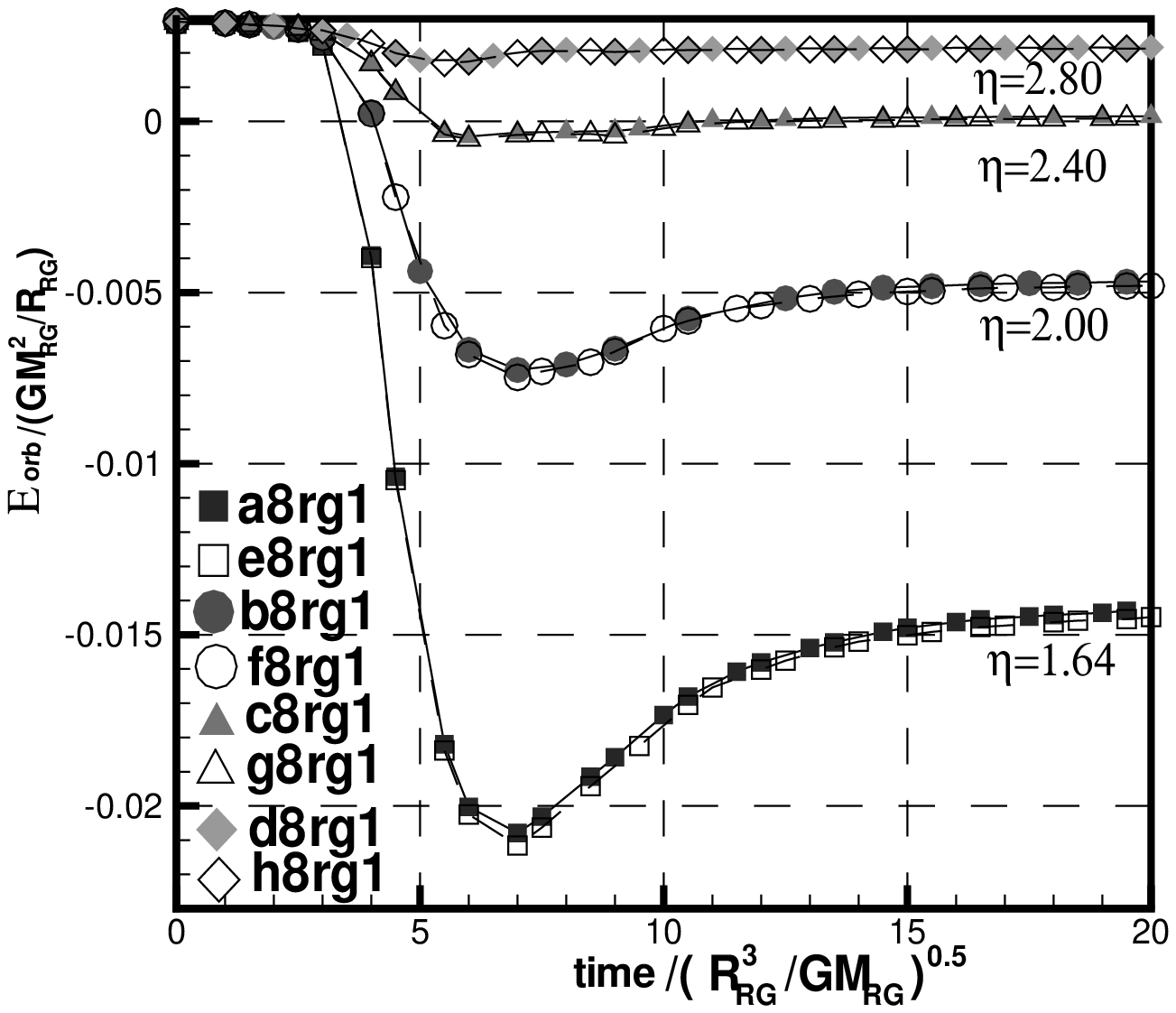}
\caption{Time variations in the orbital energy, $E_{\rm orb} = (1/2) \mu_0 v_{\infty}^2 - \Delta E$; 
    filled and open symbols denote the models of $\alpha_{\rm SPH} = 1.0$ and $\alpha_{\rm SPH} = 0.1$, respectively.  
\label{fig:timevar-e}}
\end{figure}

\begin{figure}
\plotone{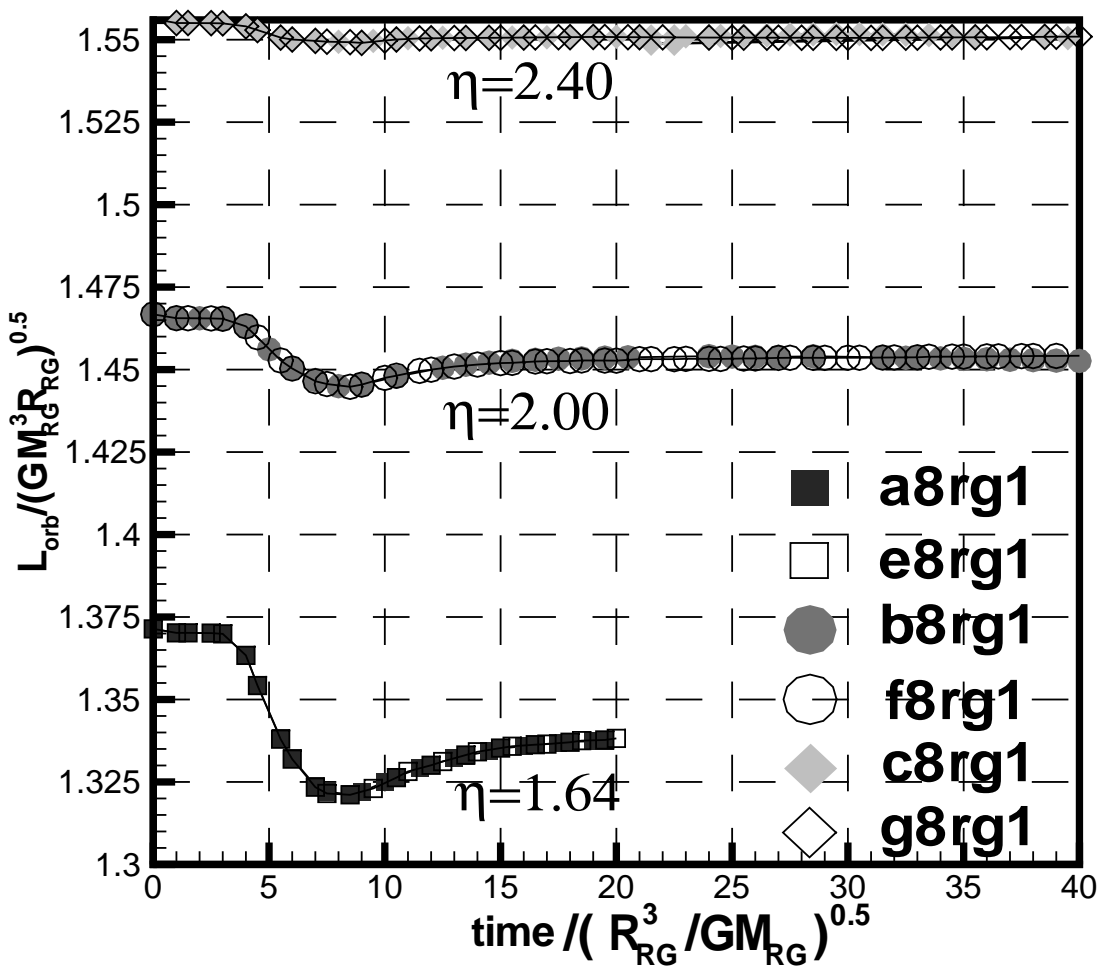} 
\caption{Time variations in the orbital angular momentum, ($L_{\rm orb} = \mu_0 b v_{\infty} - \Delta L_{\rm RG}$). 
Filled and open symbols denote the models of $\alpha_{\rm SPH} = 1.0$ and $\alpha_{\rm SPH} = 0.1$, respectively.  
\label{fig:timevar-am} }
\end{figure}

\begin{figure}
\plotone{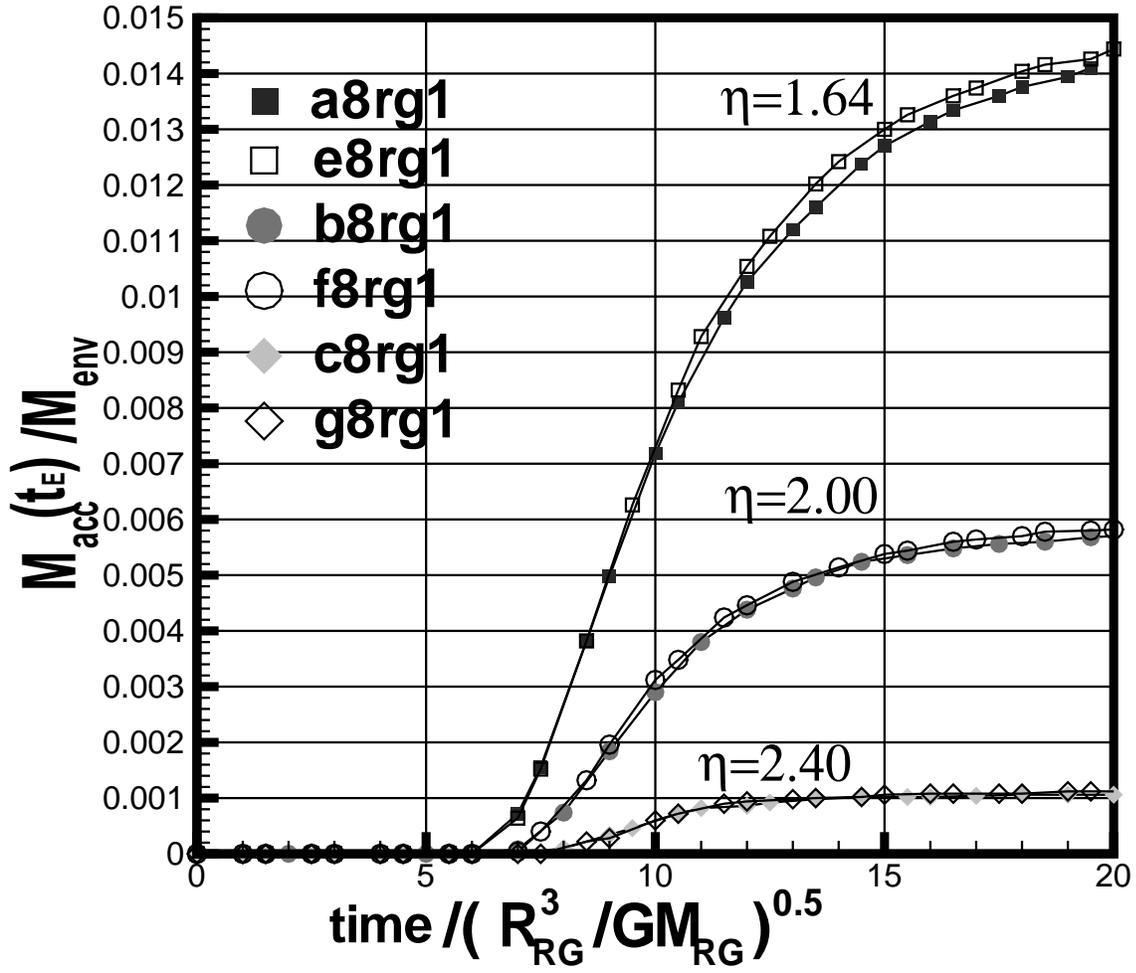}
\caption{Time variations of the accreted mass, $M_{\rm acc}$, onto the secondary point mass; 
   Symbols have the same meanings as in Fig.~\ref{fig:timevar-am}. } 
\label{fig:timevar-macc}
\end{figure}

Figures~\ref{fig:timevar-e} and \ref{fig:timevar-am} show the time variations of orbital energy and angular momentum ($E_{\rm orb} = \mu_0 v^2_\infty - \Delta E$ and $L_{\rm orb} = \mu_0 b v_\infty - \Delta L$), respectively, for the $20R_{\sun}$ models.  
   As the periastron is approached, both decrease rapidly and reach the minimum after the periastron passage. 
   Then, they turn to increase gradually to resume the loss up to $\sim 28 \%$ at the largest case, and approach to asymptotic constant values.  
   Figure~\ref{fig:timevar-macc} shows the time variation of mass, $M_{\rm acc}$, trapped by the perturber;  
   the beginning of mass accretion coincides with when the orbital energy and angular momentum hit the minimum, and the increase in the accreted mass follows the curves of the latter$^{\prime}$ recovery.    
   This indicates that the orbital energy and angular momentum once received by the surface matter are slowly returned back to the orbital motion by the accretion process.  

At the end of our simulations, the motions of two stars tend to settle in the asymptotic orbits, and the characteristics no longer change. 
   We present the values of $\Delta E (t_E)$, $\Delta L (t_E)$ and $M_{\rm acc} (t_E)$ at the end of our simulation at $t = t_E =20$ or $40 \tau_{\rm RG}$ in Tabel~\ref{tab:model=alpha}.  
   The deposited energy and angular momentum into the red giant envelope increase precipitously with decreasing periastron distance, and for $r_{\rm p} \lesssim 2 R_{\rm RG}$, both of them become appreciable in comparison with the gravitational binding energy and the angular momentum corresponding to the critical rotation of red giants, respectively. 
   The angular momentum of $\Delta L (t_E) \gtrsim 0.01 I_{\rm RG} \Omega_{\rm RG}$ (where $I_{\rm RG}$ is the moment of inertia of red giant) can be deposited, which is necessary to explain such fast rotators as observed for the horizontal branch stars in the globular clusters.  
   The accreted mass onto the main sequence stars also shows a similar tendency, amounting to be comparable with the mass in the surface convective zone of main sequence stars near the turn-off stars in the globular clusters \citep[$\sim 0.003 \msun$; see, e.g.,][]{suda06}.  
   As the perturber mass decreases by 25\% from $M_{\rm MS} = 0.8$ to $0.6 M_\odot$, the deposited energy and angular momentum decrease by a factor of $1.6 - 2.2$, and the accrete mass decreases by a slightly larger factor of $2.3 - 2.5$.  
   When the two red giant models are compared at the same encounter closeness parameter $\eta$, the interactions tend to be weaker for red giant models of larger radius, when normalized with respect to their radii, giving smaller $\Delta {E} (t_E) / (G M_{\rm RG}^2 /R_{\rm RG})$, $\Delta {L} (t_E) / (G M_{\rm RG}^3 R_{\rm RG})^{1/2}$, though the differences remain less than a factor of $ 2$.  
   This is attributable mainly to the smaller envelope mass involved in the tidal deformations (decreasing by 50 \%) for the model of the larger radius.  

In our simulations the border between the tidal capture and the fly-by, i.e., whether two stars form a bound system or fly apart after the encounter, lies in the range of $2.14 < r_p / R_{\rm RG} < 2.25$ for $20 R_\sun$ and $1.41 < r_p / R_{\rm RG} < 1.65$ for $85R_\sun$, respectively. 
   Our tidal capture limit is somewhat larger than obtained from the linear analysis by \citet{mcmillan90}, who give the range $r_p /R_{\rm RG} = 1.5 - 1.7$ for a capture of a $0.5 M_{\sun}$ dwarf (see Table~\ref{tb:capturelimit}) and \citet{bailyn1988}, who give the range $r_p /R_{\rm RG} = 1.0 - 2.0$ for a capture of a $1.4 M_{\sun}$ neutron star.
   As for the dependence on the red giants models, the case of $85R_\sun$ results in the fly-by for closer encounters than the case of $20R_\sun$ when compared between the models of the same periastron distance normalized by the stellar radius. 
   This is due to the smaller binding energy of envelope, which directly affects the capture condition $\Delta E  (t_E) > (1/2) \mu_{0} v_{\infty}^2$, as already discernible in the linear analysis by \citet{mcmillan90}.  
   On the other hand, a larger radius causes relatively greater effects on the angular momentum deposition and mass accretion;  
   for the red giants of late evolutionary stages, even the fly-by encounters can give the sufficient amounts of angular momentum and accreted mass to explain the fast rotators of HB stars and to disguise the surface of the main sequence star with accrete matter.  

\section{Parameter Dependences of Transfer Characteristics and Fitting Formulae}  

\subsection{Tidal Energy Deposition and Angular Momentum Transfer}

The linear perturbation theory has been developed by \citet{press77} and \citet{lai97} to evaluate the transfer of energy and angular momentum through the dynamical tides;  
    according to their results, the parameter dependences of these quantities are given explicitly in eqs.~(\ref{eq:lnfit-energy}) and (\ref{eq:lnfit-am}) for the $l=2$, $f$-modes (see Appendix B).  
   In order to separate the effects of the secondary mass, we may define $\Delta \tilde{E}$ and $\Delta \tilde{L}$ as 
\begin{eqnarray}
\Delta \tilde{E} & \equiv & \Delta E (t_E) / \left [(G M_{\rm RG}^2/ R_{\rm RG}) (M_{\rm MS} / (M_{\rm RG} + M_{\rm MS}))^2 \right ] \\
\Delta \tilde{L}  & \equiv & \Delta L (t_E) / \left [(M_{\rm RG} R_{\rm RG}^2 \Omega_{\rm RG}) /(M_{\rm MS} / (M_{\rm RG} + M_{\rm MS}))^2 \right ].    
\end{eqnarray}
   In the linear theory, $\Delta \tilde{E}$ and $\Delta \tilde{L}$ are expressed in terms of the transfer functions, $T_2(\eta; Q_{02}, \omega_{02})$ and $S_2(\eta; Q_{02}, \omega_{02})$, given in eq.~(\ref{eq:transferfunc1}) and eq.~(\ref{eq:transferfunc2}) in Appendix B, respectively:  
   here $\omega_{02}$ and $Q_{02}$ are the frequency and overlap integral of $l=2$, $f$-mode oscillations and for the red giants, given by interpolation formulae of eqs.~(\ref{eq:q2int}) and (\ref{eq:w2int}) in Appnendix B as a function of radius. 
   Since the $l=2$ modes dominate over the oscillatory motions even in the non-linear regime, we may utilize these transfer functions and seek the fitting formulae of $\Delta \tilde{E}$ and $\Delta \tilde{L}$ as a function of $\eta$ for the given models of red giant.  

Further, the transfer characteristics obtained by the numerical simulations differ also with the internal structure of red giant models, as seen above.  
   In the linear theory, the dependences of the internal structure are included in the transfer functions, in particular, through the overlap integral $Q_{02}$. 
   In the tidal interactions via the torque, the coupling is given by the moment of inertia of the envelope since the core acts as an inert source of gravity. 
  As a corollary, the transfer functions can be scaled with the inertia of envelope.   
   We define the non-dimensional moment of inertia, $\tilde{I}$, as
\begin{equation}
\tilde{I} \equiv \int 4 \pi r^4 \rho d r / M_{\rm RG} R_{\rm RG}^2 \simeq  0.15 (M_{\rm env} / M_{\rm RG}), 
\label{eq:tildeI}
\end{equation} 
   for the red giant models. 
   The approximation in the rightmost member in eq.~(\ref{eq:tildeI}) follows from the similarity of the envelope structure in the red giants when normalized with respect to the surface radius and the envelope mass, as shown in Fig.~\ref{fig:redgiant}.  
   For a main sequence model of polytrope $N=1.5$, we have $\tilde{I} \equiv \int 4 \pi r^4 \rho d r / M_{\rm MS} R_{\rm MS}^2 \simeq 0.2$.  

\begin{figure}
\epsscale{0.8}
\plotone{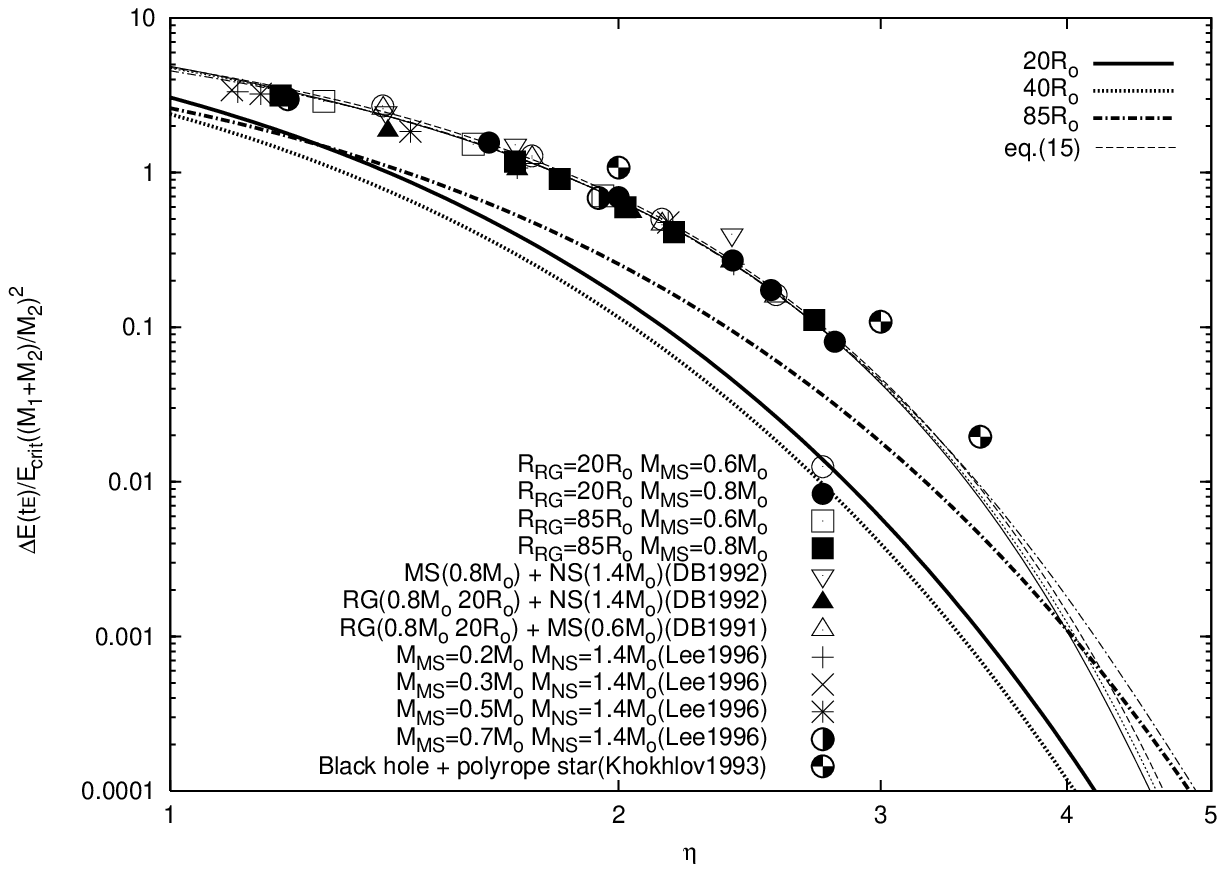}
\plotone{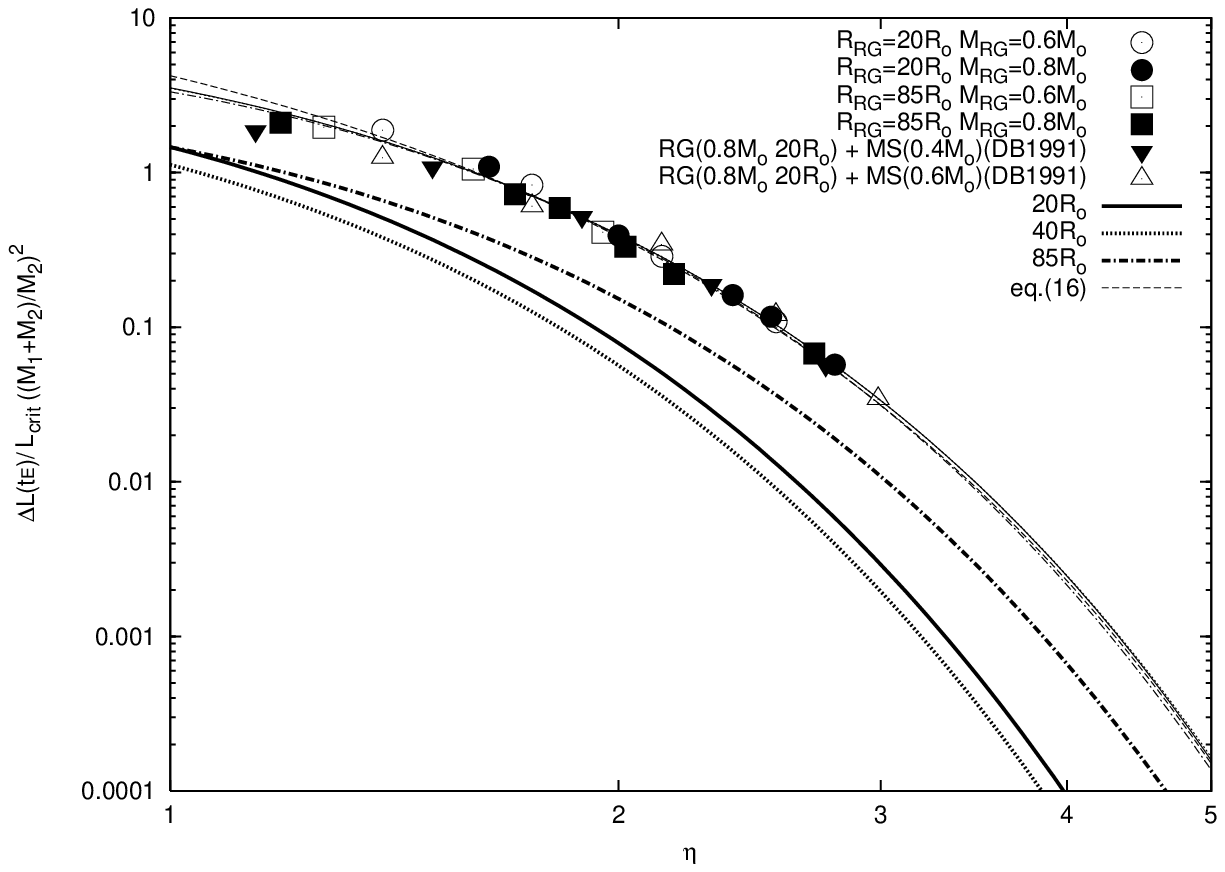}
\caption{
The deposited energy and angular momentum, normalized by the inertia, $\Delta \tilde{E}/\frac{1}{2}\tilde{I}$ (top panel) and $\Delta \tilde{L}/\tilde{I}$ (bottom panel) from the orbital to stellar internal motions are plotted as a function of the encounter closeness parameter $\eta$ for the red giant model of radius, 20 and $85 R_\odot$ (circles and squares) with the main sequence perturber of mass, 0.6 and $0.8 \msun$ (open and filled symbols). 
   Also plotted are the results by other authors: 
   a red giant of mass $0.8 \msun$ and radius $20 R_\odot$ with a main sequence of mass, $0.4 \msun$ and $0.6 \msun$, and with a neutron star of mass $1.4 \msun$ by \citet{davies91,davies92} (inverted filled, open and filled triangles):   
   a $0.8 \msun$ main sequence with a neutron star of mass $1.4 \msun$, by \citet{davies92} (inverted open triangles):  
   a $0.2 \msun$, $0.3 \msun$, $0.5 \msun$ and $0.7 \msun$ main sequence star with a $1.4 \msun$ neutron star, by \citet{lee96} (plus signs, crosses, asterisks and waning moons): 
   a polytrope with a black hole by \citep{khokhlov93a, khokhlov93b} (windmills).  
   Thin solid, dotted and dash-dotted lines denote the fitting curves given in eq.~(\ref{eq:Ecrit}) and eq.~(\ref{eq:Lcrit}) for $20 R_\sun$, $40R_\sun$ and $85R_\sun$, respectively and the dashed line denots eq.~(\ref{eq:delEfit-unique}) and eq.~(\ref{eq:AMfit-unique}), while thick lines denote the corresponding results derived from the linear theory.
   See text for details. 
\label{fig:tildeoinertia} }
\end{figure}

Figure~\ref{fig:tildeoinertia} shows $\Delta \tilde{E} / (\tilde{I} / 2)$ and $\Delta \tilde{L} / \tilde{I}$ as a function of $\eta$ for the models with the different red giant models of radii 20 and $85 R_\odot$ (denoted by circles and squares, respectively) and with the different main sequence stars of mass 0.6 and $0.8 \msun$ (denoted by open and filled symbols, respectively). 
   They form a single curve on each panel, independent not only of the perturber mass but also of the red giant models.  
   The models computed by \citet{davies91} and \citet{davies92} are also plotted in this figure;
   their models of a $0.8 \msun$ red giant and $20 R_\odot$ give very good agreement with ours for the encounter not only with a $0.6 \msun$ main sequence star (open triangles) but also with a $1.4 \msun$ neutron star (filled triangles). 
   The encounter with neutron stars result in slightly smaller energy deposition (about several tenths) for close encounters of $\eta \lesssim 2$, which is attributable to larger accretion radius of neutron star since the accretion of larger mass onto the perturber returns a larger portion of energy from the outer elongated part of red giant envelope to the orbital motion.  
   As for the angular momentum, the deposition is slightly smaller in their red giant models at close encounters of $\eta < 2$ than in ours, which may stem from the difference in the criterion of mass loss particles, giving a larger mass loss to their models, or from larger radii of red giant, used in the normalization, with taking into account the swell of the red giants during the encounter. 

In the upper panel, we also plot the results for the encounter simulations of main sequence stars with a neutron star and with a black hole by \citet{davies92} and \citet{lee96} and for the encounter of a $N=1.5$ main sequence star with a black hole by \citet{khokhlov93a}.  
   We approximate the main sequence to a polytrope of index $N=1.5$ and take $\tilde{I} = 0.2$.  
   These models fall very closely along the same curve as our models, and seemingly compose a single group despite the difference not only in the mass of the main sequence stars but also in the mass ratio.  
   The models by \citet{lee96} give slightly smaller values than those by \citet{davies92}, which may stem from the different criteria and treatment of the particles that accrete onto the neutron star. 
   The models by \citet{khokhlov93a} show slightly larger than those by \citet{davies92}, which may stem from the neglect of the accretion effect. 

This convergence may be related to the fact that the ratio, $\eta \omega_{02}$, between the timescale of periastron passage and the timescale of envelope oscillations decreases near to the unity in the non-linear regime for small $\eta$.  
   The similar tendency that the dependence on the stellar models becomes weaker for smaller $\eta$ is also discernible in eqs.~(\ref{eq:lnfit-energy}) and (\ref{eq:lnfit-am}) from the linear perturbation theory.   
   These results for the linear regime are also plotted in this figure, and the comparison with the results of non-linear regime indicates that the latter effect enlarges the deposition of energy and angular momentum by a factor of several and up to ten, while it becomes saturated and slightly dwindles for smaller $\eta < 2$ because of the mass accretion onto the perturber. 

We may take advantage of the convergence in the non-linear regime to evaluate the transfer characteristics for other red giant models and to seek the fitting formulae that express $\Delta E (t_E) $ and $\Delta L (t_E) $ in terms of model parameters. 
   We define the critical rotation energy and angular momentum of model stars as $E_{\rm crit} = (1/2) \tilde{I} (G M_{\rm RG}^2 / R_{\rm RG})$ and $L_{\rm crit} = \tilde{I} M_{\rm RG} R_{\rm RG}^2 \Omega_{\rm RG}^2$, respectively, with the dependence on the moment of inertia taken into account, and assume the following fitting formulae that converge to the results of linear theory for distant encounters;
\begin{eqnarray}
\Delta E (t_E) / [E_{\rm crit} \{ M_{\rm P}/(M_{\rm RG}+M_{\rm P}) \}^2] & = &  (2 /\tilde{I}) \eta^{- 4} T_2(\eta; Q_{02} , \omega_{02}) [1 + \exp(a_{\rm 1}  \eta^2 + b_{\rm 1}  \eta  + c_{\rm 1})], \label{eq:Ecrit}  \\
\Delta L (t_E) / [ L_{crit} \{ M_{\rm P}/(M_{\rm RG}+M_{\rm P}) \}^2 ] & = & (1/\tilde{I})  \eta^{- 4} S_2(\eta , Q_{02} , \omega_{02}) [1 + \exp(a_{\rm 2} \eta^2 +  b_{\rm 2} \eta + c_{\rm 2})]. \label{eq:Lcrit}
\end{eqnarray} 
   where $M_{\rm P}$ is the perturber mass.   
   For the red giant models of various evolutionary stages, the transfer functions are computed with the estimates of $\omega_{02}$ and $Q_{02}$ from eqs.~(\ref{eq:w2int}) and (\ref{eq:q2int}) in Appendix B. 
   For the main sequence stars, the values of $\omega_{02}$ and $Q_{02}$ are taken from \citet{lee86} for a $N=1.5$ polytrope.   
   We may determine the coefficients in these formulae by applying the fitting procedure with the non-linear least-square Marquardt-Levenberg algorithm to our data plotted in Fig.~\ref{fig:tildeoinertia} for the red giant models of radii 20 and $85 R_\odot$.   
   The fitting curves are plotted in the figure, which converges into a unique relationship in the non-linear regime for the red giant models of different radii.   
   The fitting curves for the models with other radii (core masses) also derived by adopting the same data in the non-linear regime, as shown for the model of $40R_\odot$ in the figure.      
   These fitting parameters, thus obtained for the various red giant models, are expressed as the second order polynomials of radius as: 
\begin{eqnarray}
a_{\rm 1} & = & -\hbox{7.8 E-05} (R_{\rm RG}/R_\sun)^2 + \hbox{9.6 E-03}(R_{\rm RG}/R_\sun) - 0.66 \nonumber \\
b_{\rm 1} & = & \hbox{9.0 E-05} (R_{\rm RG}/R_\sun)^2 - 0.028 (R_{\rm RG}/R_\sun) + 3.7  \nonumber \\
c_{\rm 1} & = & -\hbox{5.4 E-04} (R_{\rm RG}/R_\sun)^2 + 0.077 (R_{\rm RG}/R_\sun)  -4.6  \label{eq:cfit1} \\
a_{\rm 2} & = & -\hbox{4.6 E-06} (R_{\rm RG}/R_\sun)^2 + 0.00104 (R_{\rm RG}/R_\sun)  -0.05945 \nonumber \\
b_{\rm 2} & = & \hbox{2.6 E-04} (R_{\rm RG}/R_\sun)^2 + 0.0127 (R_{\rm RG}/R_\sun) + 1.02 \nonumber \\
c_{\rm 2} & = & \hbox{2.2 E-04} (R_{\rm RG}/R_\sun)^2 + 0.036 (R_{\rm RG}/R_\sun) - 1.42 \nonumber
\end{eqnarray}  

Furthermore, in the non-linear regime, the deposited energy and angular momentum during the encounters may be given simply as the functions of $\eta$ in the form:  
\begin{eqnarray}
\Delta {E} (t_E) & = & E_{\rm crit} (M_{\rm P} / \{M_{\rm RG} + M_{\rm P}\})^2 \exp (2.718 -0.761 \eta - 0.386 \eta^2), \label{eq:delEfit-unique} \\
\Delta {L} (t_E) & = &  L_{\rm crit} (M_{\rm P} /\{ M_{\rm RG} + M_{\rm P}\})^2   \exp (3. 735 - 2.237 \eta - 0. 040 \eta^2) ,     
\label{eq:AMfit-unique}
\end{eqnarray} 
   which are plotted in this figure by broken lines. 
   Note that they coincide with the expressions obtained above in eqs.~(\ref{eq:Ecrit}) and (\ref{eq:Lcrit}) for close encounters of $\eta \lesssim 4$ and $\eta \lesssim 10$,  respectively.  
   These formulae are applicable not only to the encounter of red giants but also to that of main sequence stars of arbitrary mass and radius as long as the stellar mass is small enough for the surface convection to develop deep enough to be approximated by a polytrope of $N = 1.5$.  

\subsection{Accreted Mass onto Main Sequence Stars}

During the encounter, matter near the very surface may gain a lot of energy and angular momentum from the orbital motion to expand and eventually be captured by the perturber.
   The matter accreted onto the perturber returns the acquisitions back to the orbital motion, which may reduce the transfer of energy and angular momentum at small $\eta < 2$, as stated above. 
   As seen from Table~\ref{tab:model=alpha}, the accreted mass, $M_{\rm acc}(t_E)$, turns out to be nearly the same among the models of the similar values of $\Delta {E} (t_E) / (GM_{\rm RG}^2 / R_{\rm RG})$ regardless of the radius of red giants.  
   This is indicative that the amount of accreted mass is related to the deposited energy.   
   In this section, we study the relationship between the accreted mass and other physical quantities and attempt to express the accreted mass as a function of the model parameters. 

In the case of a star filling the Roche lobe in a close binary, \citet{paczynski72} argue that the mass transfer rate is related to the excess, $\Delta R$, of stellar radius over the Roche lobe under the assumption of polytrope;  
    the principal part of parameter dependences of the transfer rate (see their eq.~[A21] ) is approximated by; 
\begin{eqnarray}
\dot{M} \sim 4 \pi A^2 \left( {G M \over A} \right)^{N+0.5} \left( {M_{\rm RG} \over M } \right) ^{N + 1.5} K^{-N} \left( {\Delta R \over R_{\rm L}}\right)^{N+ 1.5},
\end{eqnarray}
   where $A$ and $M$ are the separation and total mass of the binary system, respectively.  
   Here we neglect the weak dependence on the mass ratio, and in particular, take the Roche radius $R_{\rm L} / A \approx 0.38$ (cf. eq.~[\ref{eq:Roche-lobe}]).  
   Although the flow is not in steady state in our case, the timescale of flow through the inner Lagrangian point is slower than the dynamical timescale of stellar envelope, and hence, we may assume the same dependences for the accreted mass. 
   Further since the orbit is eccentric and not circular, it is difficult to estimate ${\Delta R} / {R_{\rm L}}$ exactly.  
   And yet, it seems natural to $\Delta R / R_{\rm L}$ to be related to the deposited energy and we may well assume the following relation; 
\begin{equation}
\frac{(d \phi_L / d r) \Delta R_{\rm L}}{GM / R_{\rm RG}} = \frac{\Delta R_{\rm L}} {R_{\rm L}} \frac{R_{\rm RG}}{R_{\rm L}} \sim f ( \Delta E / [G M_{\rm RG}^2 / R_{\rm RG}]) 
\end{equation}
   with taking account of the work against the gravitational potential ($\phi_{\rm L}$) at the Roche lobe surface.  
   By using the relation between the polytropic constant $K$ and the stellar surface characteristics in eq.~(\ref{eq:redgiantK}), we then have;  
\begin{equation}
\dot{M} \sim M_{\rm RG} (\frac{M_{\rm RG}}{M_{\rm env}})^{N-1}  \frac{\Omega_{\rm RG}^{3}}{{\Omega_{\rm pass}}^2} f ( \Delta E / [G M_{\rm RG}^2 / R_{\rm RG}])^{N+1.5}. 
\label{eq:dotM-enc}
\end{equation}
    Here we have replaced the orbital angular velocity by the instantaneous angular velocity, $\Omega_{\rm pass} (= \sqrt{G M / A^3})$, of circular orbit at periastron distance.  
    Consequently, multiplying eq.~(\ref{eq:dotM-enc}) by the periastron passage time $ \sim \Omega_{\rm pass}^{-1}$ and putting $N=1.5$ lead us to:  
\begin{equation}
M_{\rm acc} \sim M_{\rm RG} ({M_{\rm env} / {M_{\rm RG}}})^{-1/2} \eta ^3 f ( \Delta {E} / [GM_{\rm RG}^2/ R_{\rm RG}])^{N+1.5}. 
\label{eq:simeq-macc}
\end{equation}

\begin{figure}
\plotone{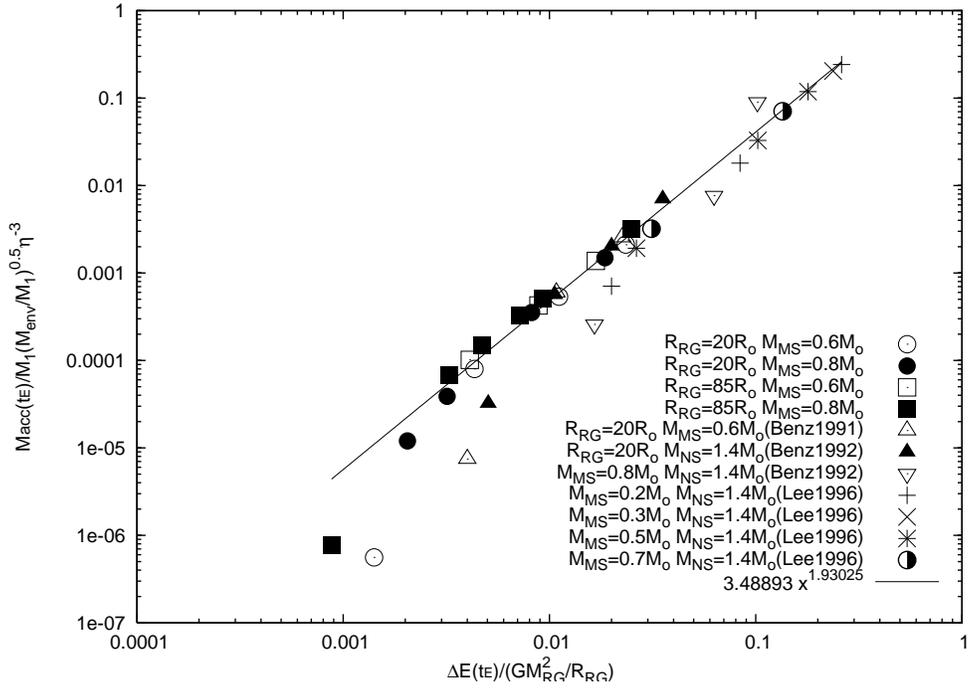}
\caption{The accreted mass, multiplied by the root of the mass fraction of envelope of red giant and divided by third power of encounter closeness parameter, $\eta$, as a function of the deposited energy, $\Delta E (t_E)$, into the red giant envelope. 
   Symbols are the same as Figure~\ref{fig:tildeoinertia} and solid line represents a power-law fitting.  
   See text for details.
\label{fig:macc-delE}}
\end{figure}

Figure~\ref{fig:macc-delE} shows the accreted mass, $M_{\rm acc} (t_E) $, divided by $M_{\rm RG} ({M_{\rm env}} / {M_{\rm RG}})^{-1/2} \eta^{3}$, against $\Delta {E} (t_E) / ( GM_{\rm RG}^2 / R_{\rm RG})$.  
   It is clearly seen that our numerical results, denoted by open and filled circles and squares, form a unique relationship, indifferent of the mass of perturber and of the red giant model; 
   a power-low relation is discernible in the range of $\Delta {E} (t_E) > 3 \times 10^{-3} GM_{\rm RG}^2 / R_{\rm RG}$.  
   For smaller $\Delta {E} (t_E)$, the accreted mass tends to drop off from the power-law relationship because of the low mass resolution due to the limited number of gas particles;  
   in actuality, only a few gas particles are accreted around $\Delta {E}(t_E) \sim 3 \times10^{-3} GM_{\rm RG}^2 / R_{\rm RG}$.    
   In this figure, we plot the results of the simulations by \citet{davies91}, \citet{davies92}, and \citet{lee96}, and find that their values also fall very closely onto the same relationship for relatively large deposited energy, while the lower mass resolution in these earlier simulations (7500 and 9185 SPH particles, respectively) causes the deviation at larger $\Delta {E}(t_E) / GM_{\rm RG}^2 / R_{\rm RG}$.   
   If these results for the encounter of main sequence stars are included, the power relationship holds in the range of mass accretion over four orders of magnitude or more.  

\begin{figure}
\plotone{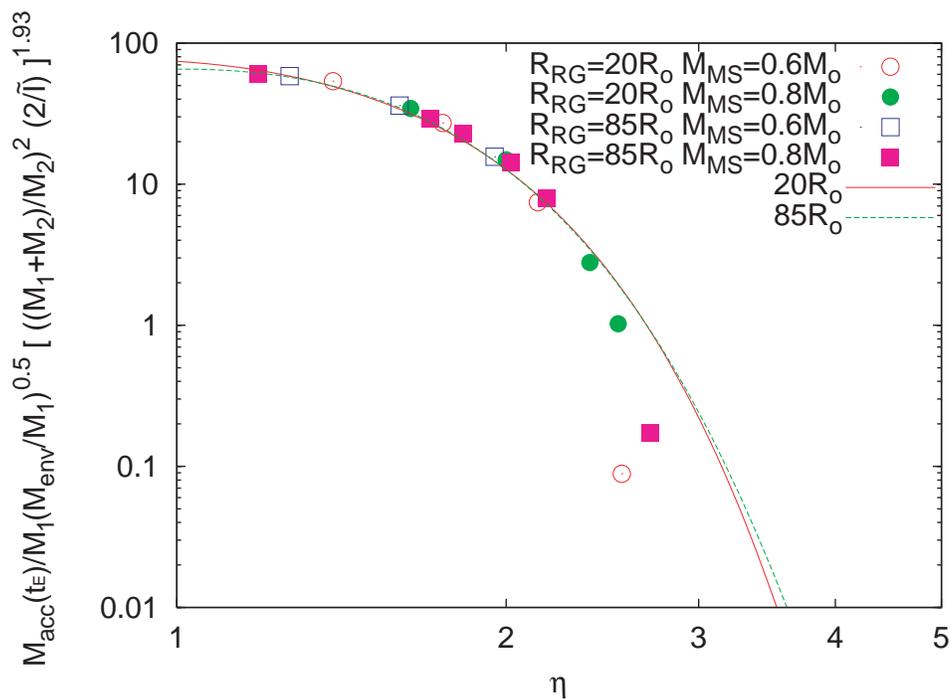}
\caption{The accreted mass, multiplied by $(M_{\rm env}/M_1)^{1/2} [2 / \tilde{I} [ (M_1  M_2)/M_2 ]^2 ]^{1.93}$, as a function of $\eta$.    
   Symbols are same as Figure \ref{fig:macc-delE}. 
   Solid and dashed curve are derived from {eq.~(\ref{eq:fitting-macc})} and the fitting formula eq.~(\ref{eq:Ecrit}) for the red giants models of $20R_\sun$ and $85R_\sun$, respectively (see text for details).}
\label{fig:macc-tildaE}
\end{figure}

The power-law fitting to eq.~(\ref{eq:simeq-macc}) yields
\begin{equation}
    M_{\rm acc} / M_{\rm RG} = 3.5 ({M_{\rm env}}/{M_{\rm RG}})^{-1/2} \eta^{3}  [ \Delta {E} (t_E) / (GM_{\rm RG}^2 / R_{\rm RG})]^{1.93},   
\label{eq:fitting-macc}
\end{equation}
   which gives the accreted mass as a function of model parameters along with the fitting formula of $\Delta {E} (t_E)$ in eq.~(\ref {eq:Ecrit}) or in eq.~(\ref{eq:delEfit-unique}).  
   This relation implies that $\Delta R / R_L \propto \Delta E^{1.93/3}$.  
   We show the comparison between this relation and the results of simulations in Figure~\ref{fig:macc-tildaE} as a function of $\eta$ which is more useful than that of $\Delta {E} (t_E)$.  
   It shows a good agreement for the both red giant models.  
   Since we may regard the deviations for large $\eta (\gtrsim 2.5$) as due to the low resolution in mass in the simulations, this gives a reasonable fitting for the accreted mass as a function of $\eta$ for any given set of the model parameters of encounters.  

\section{CONCLUSIONS AND DISSCUSION}

We have performed the SPH simulations of tidal encounter of red giants with environment stars and investigate the characteristics of stellar interactions for a variety of sets of parameters, the evolutionary stages of red giant, the mass of perturber stars and the assumed strength of viscosity as well as the orbital parameters of encounter.  
   Based on our results and the other extent models, we discuss the dependences of interactions in the non-linear regime on the stellar and encounter parameters, and proposed formulae to describe the energy and angular momentum deposition to red giants, and the mass accretion onto the perturber stars in simple and convenient forms as a function of these parameters. 
  Our main quantitative results are as follows; 


\begin{enumerate}
\item 
   We obtain the both amounts of energy and angular momentum, transferred from the orbital motion into the oscillation and rotation of red giants, during tidal encounters by numerical simulations. 
   The angular momentum deposited in the red giants can be large enough to rotate the envelope at rate $\Omega \gtrsim 0.01 \Omega_{\rm RG}$ for the encounter of periastron distance $r_p / R_{\rm RG} \lesssim 2.5$ (or the impact parameter $b / R_{\rm RG} \lesssim 15.7$) and hence for such encounters that ended in the two stars flying apart.  
   For still closer encounters, it increases to give the rotation rates significantly exceeding $\Omega \simeq 0.1 \Omega_{\rm RG}$. 
   Accordingly, the tidal encounter works as the source of angular momentum necessary to trigger rotational mixing in the red giants and also to explain the origin of fast rotators observed among the horizontal branch stars.  
   For larger radius of red giant model, and hence, later stage of evolution, the transferred angular momentum increases while the energy deposition decreases since the transferred quantities are scaled with the stellar parameters.  
   The fitting formulae are derived to describe these quantities as a function of the mass and radius of red giants, subject to the perturbation, the mass of main sequence stars as the perturber, and the impact parameter.  
   Further we show that these transferred quantities, when normalized with respect to the momentum of inertia of models, are given solely as functions of the encounter closeness parameter $\eta$,  
   Our fitting formulae agree well with the results of encounter simulations by other authors, and can even reproduce the results for the encounters of main sequence stars approximated to a polytrope of index $N = 1.5$.  

\item  
    With aid of fine mass resolution, we demonstrate that the main sequence stars can capture gas from the red giant envelope sufficiently to disguise their surface with accreted matter even for the encounters that ended in fly-by with an evolved red giant. 
    The accreted mass onto the main sequence as a perturber during the tidal encounters is shown to be in a direct relationship with the energy deposition into the red giants.  
   We also derive the formula, which predicts the accreted mass as a function of impact parameters for given stellar parameters and are applicable to the encounters involving not only the red giants but also the main sequence stars.  
\end{enumerate}

The derived formulae are useful in determining the periastron distance of the tidal capture limit for the encounter of various model parameters.  
   They also be useful in inquiring whether some stellar objects in the globular clusters, for example the red giants and the main sequence stars with abundance anomalies and the fast rotating horizontal brunch stars, which cannot be explained through the framework of the normal stellar evolution, can be produced through the stellar interactions.  
   In our computations, the mass accretion rate may exceed the Eddington limit (${\dot M}_{\rm Edd} = 4 \pi c R_{\rm MS} / \kappa_e$) on the surface of main sequence star for very close encounters, but since it remains below the Eddington limit at the accretion radius, we may well assume that the accreted mass mostly settles on the surface of main sequence stars with loosing thermal energy.  
   In the following, we discuss the application of these formulae and the possibility that such stellar objects have their origins in the stellar encounters.


\subsection{Tidal Capture Limit and Comparisons with Other Works}

In our simulations, the tidal capture limits, $r_{p, \rm cap}$ in periastron distance and $\eta_{\rm cap}$, in $\eta$, are estimated from the condition that $\Delta E (t_E) = (1/2) \mu v_\infty (10 \hbox{ km s})^2$ with use of eq.~(\ref{eq:Ecrit}) or eq.~(\ref{eq:delEfit-unique}). 
   Our estimates are given in Table~\ref{tb:capturelimit} for the two red giant models with the perturber masses of $0.6 M_{\sun}$, $0.8 M_{\sun}$ and $1.4 M_{\sun}$. 
   For the red giant of larger radius, the tidal capture limits decrease slightly ($\sim 10 \% $) when normalized with respect to the radius of red giant, but increase in the physical dimensions nearly in proportion to the surface radius;  
   they slightly increase with the mass of perturber. 
   For the encounter of $M_{\rm RG} = 0.8M_{\sun}$ red giant with a $M_{\rm MS}=0.6M_{\sun}$ main sequence star, \citet{davies91,davies92} give the periastron distances for the tidal capture limit in the range of $2.00 < r_{p, \rm cap} / R_{\rm RG} < 2.25$, and our estimate, $r_{p, \rm cap} / R_{\rm RG}= 2.1$, as seen from Table~\ref{tb:capturelimit}, resides in their range.   
   On the other hand, our estimates turn out to be larger by $\sim 20\%$ than those obtained from the linear analysis by \citet{mcmillan90}, as listed in the table.  

On the other hand, \citet{khokhlov93a} and \citet{khokhlov93b} compute the encounter of a polytrope star of mass $M_{*}=0.8M_{\sun}$ with a black hole of mass $M_B \gg M_*$ for various values of polytropic index (the relative velocity is 100 km/s at infinity).  
   The tidal capture limit decreases from $\eta_{\rm cap}= 2.75$ for $N = 1.5$ to 2.20 and 1.55 for $N = 2.0$ and 3.0, respectively.  
   This demonstrates that for larger polytrope index, the energy deposition rate of the star becomes smaller, because of the increase in the mass concentration toward the center and of decrease in the moment of inertia for a given mass and radius.  

\subsection{Relevance to the Origin of Stars with Anomalous Abundances in Globular Clusters}

We may apply our fitting formulae to examine the possibility that the abundance anomalies observed both for red giants and main sequence stars in some globular clusters can be explained in terms of the stellar interactions during the close encounters.  
  As the origin of these objects, we propose the following scenario; 
   1) The anomalies of red giants are generated through the flash-assisted deep mixing mechanism that is triggered by the injection of angular momentum into their envelope during the encounter with environment stars \citep{fujimoto99}: 
   2) The main sequence stars gain the abundance anomalies as a result of the surface pollution by accreting matter from the red giants which have already developed these anomalies.  
  We evaluate whether these work under the conditions prevailing in globular clusters.  

First as for the point 1), \citet{fujimoto99} argue that the angular momentum of $\Delta L_{\rm RG} / L_{\rm crit} \gtrsim 1/100$ is necessary to induce the flash-assisted deep mixing, and the transfer of angular momentum of this order occurs during the encounter of $\eta \lesssim 3$, and hence, $r_p \sim 2.6 R_{\rm RG}$ from our formulae.  
   The transfer of angular momentum may have relevance to the a bimodal distribution of rotation velocity that horizontal branch stars display with the fastest rotators distributed on the cooler (redder) side of the branch whereas with the slower rotators spread over wider range on the branch.  
   \citet{suda06} suggest that the different modes of helium mixing mechanism may result according to when the stars undergo the close encounter and the deposition of angular momentum on the RGB and influence the horizontal branch morphology;  
   the injection of angular momentum may invoke hydrodynamical instabilities due to differential rotation and invokes turbulent mixing to trigger the hydrogen-flash driven deep mixing, and the resultant helium enrichment in the envelope accelerates the evolution of RGB. 
   The stars that experience the helium enrichment at an earlier epoch on the RGB have a smaller mass of helium core and hence, located on redder side of HB and those with later mixing epoch shifts to blue-ward:  
   on the other hand, if the stars experience the close encounter near the tip of RGB, the helium-flash driven mixing, rather than the hydrogen-flash driven deep mixing, takes places and causes the largest decrease of helium core so that the stars are situated at the reddest end on the HB.  
   If the stars experience close encounter at an earlier stage of RGB, then, they become slow rotators, due to angular momentum loss through mass loss on the RGB, and hence, settle on redder side of horizontal branch, and the stars, if experiencing it at later stage of RGB, become faster rotators and settle on bluer side.  
   Finally, the stars, which undergo close encounters very close to the tip of RGB, become the fastest HB rotators, located on redder-most side of the branch.  
   The close encounters at $\eta \gtrsim 3$ can explain the fastest rotation rates observed from HB stars of $\Omega \sim 0.1 \Omega_{\rm K}$ if the angular momentum is conserved during the contraction from the RGB to the HB. 
   While the pristine angular momentum is lost effectively for such low mass stars, such fast rotation as observed for HB stars may be expected also from the synchronization of red giants in the binary systems of separations (more than a several au), and yet, it is difficult for such binaries to survive without suffering encounters in the dense stellar environment of globular clusters (see below eq.~[\ref{eq:tau-enc}]).  
   In other word, we may take the existence of HB stars of these fastest rotations as an evidence that such close encounters as $\eta \lesssim 3$ take place in these clusters.

Next as for the point 2), since the mass in the surface convective zone is of $\sim 3 \times 10^{-3} M_\odot$, the accreted mass of an order of $\sim 10^{-3} \msun$ suffices to disguise the surface abundances with those transferred from the red giant envelope with anomalous abundances.  
   From the present results, it is possible to estimate the range of periastron distance, $\eta$, that can allow the accreted mass of this order at $\eta =1.8 \sim 2.2$ and $1.5 \sim 2.0$, which correspond to $r_p = 1.9 \sim 2.1 R_{\rm RG}$ and $1.7 \sim 2.0 R_{\rm RG}$, for $R_{\rm RG} = 20 R_\odot$ and $85 R_\odot$, respectively, for masses range of main sequence stars of $0.6 M_\odot$ and $0.8 M_\odot$.  

The timescale of tidal interactions in the environment where the stellar density is $n_f \hbox{ pc}^{-3}$ and the velocity dispersion $v_{\ \infty}$, is estimated (with the gravitational focusing taken into account, because of low velocity dispersion of environment stars in the core) at: 
\begin{eqnarray}
\tau_{\rm enc} \sim 7.4 \times 10^{9} (\frac{10^{4} \hbox{ pc}^{-3}}{n_f}) (\frac{100R_{\sun}} {r_p}) (\frac{v_{ \infty}}{10 \hbox{km s}^{-1}})(\frac{M_1+M_2}{2M_{\sun}} )^{-1} {\rm yr}
\label{eq:tau-enc}
\end{eqnarray} 
   On this basis, the timescales of tidal encounters, which can bring about the mass accretion of a order of $10^{-3}M_{\sun}$ and the angular momentum transfer of $\Delta L_{\rm RG} / L_{\rm crit} \gtrsim 1/100$, are estimated at $\sim 1.5 \cdot 10^{10}$ yr for $20 R_{\sun}$ and $\sim 3 \cdot 10^{9}$ yr for $85R_{\sun}$ in the environment of $n_f = 10^{4} \hbox{ pc}^{-3}$ and $v_{\infty} = 10 \hbox{ km s}^{-1}$.  
   These time-scales seem to be too large to explain the observed these objects by tidal interactions as compared with the lifetimes on the corresponding stages of red giants, $\sim 10^8$ yr for $20R_\sun$  and $\sim 10^7$ yr for $85R_\sun$, and the accumulated number of close encounters, attendant with the abundance anomalies, seems to be no more than a few, too small to explain the observations even in rough estimates.  
   In order for the encounters to be viable, these time-scales have to be shorter more than an order of magnitude, and accordingly, there need some mechanism(s) to enhance the frequency of tidal encounter in globular clusters.  

\citet{sugimoto96} points out the importance of mass segregation in modeling dynamical evolution of star clusters to explain the observable number of mill-second pulsars in 47 Tuc; 
  without the mass segregation, theoretical estimation indicates that the formation probability of binary with a neutron star is higher in $\omega$ Cen having no collapsed core than 47 Tuc having collapsed core, although in actuality, the former has not been reported to contain any pulsars.  
   Moreover, mass segregation promotes the core collapse and make it more rapidly than in the case of single-mass component.  
   \citet{portegies01} confirm that mass segregation enriches the core of star cluster in giants and white dwarfs by N-body simulations for an open star cluster.  
  In order to explain the observable number of giants with abundance anomalies, about a half of total number of giants have to experience tidal encounters with field stars.  
  It is necessary to see whether the mass segregation can gather most of giants, though not all, in the core and can increase the two-body encounter rate by more than an order of magnitude. 
   In addition to the mass segregation, the gravo-thermal oscillations of star clusters, which has been proposed by \citet{bettwieser84} and confirmed by N-body simulation \citet{makino96}, may influence the rate;  
   since the interactions are expected to occur in the evolutional stage near the high density peak, it is necessary to investigate in a correct manner how deeply and how long such a high density state reaches and lasts to affect the encounter rates. 

For the encounter with the red giants of later stages, the pollution of main sequence stars are possible even when the two stars fly apart after the encounters.  
   For the encounter that ends in the formation of binary, it depends on the fate of two stars whether the abundance anomalies imprinted onto the companions are observable or not, and it is necessary to pursue the details of evolution of tidally captured binaries. 
   There are many effects to influence the binary evolution, such as the spin-up and mode damping rates of red giant after the encounter,  the mass transfer to the companion at subsequent periastron passages \citep[ex.][]{sep07}, the mass loss from the bound system, and also the expansion of red giant as it ascends the red giant branch.  
%
In addition, we should also take into account the tidal interactions of formed binaries with the environment stars. 
   It is conceivable as the destination of tidally captured binary that either one component is liberated through the exchange encounter with a third body, or the two stars eventually coalesce as a result of Roche lobe overflow. 
   The possibility of the exchange event depends on the evolution timescale of giants and the encounter timescale; 
   since the exchange encounter rate is proportional nearly to a semi-major axis of a binary under the assumption of point mass limit \citep{ heggie96}, then the former is expected to occur more frequently.  
   Accordingly, if tidal capture of two body encounters can contribute the modification of stellar populations, then, the exchange encounters follow at larger rates and make greater contributions.  
  This increase in the encounter rates with red giant in the binary may affect the statistics of the main sequence stars with the abundance anomalies.  
  Further, through the exchange encounters, the red giants and horizontal branch stars, now losing the mass due to the mass loss and becoming lighter than the heaviest main sequence stars, can be ejected from the binary systems to fly apart as single stars.  

In summary, the enhancement of tidal interactions, necessary to explain the observed abundance anomalies, is expected to be provided by the formation of high density core due to the gravo-thermal oscillations and by the mass segregations which enlarge the fractions of stars of the upper mass end in the core. 
   The proper understandings of these effects wait for N-body simulations of globular clusters with the stars of multi-mass spectra taken into account since star-star interactions as studied in the present work will play a critical role.  
   In these studies of dynamical evolution of stellar systems, the formulae for the transfer of energy, angular momentum, and accreted mass derived in the present work serve the purpose for incorporation these effects into the simulations.   
   It is also necessary to pursue the binary evolution and accurately explore the fate of red giants and subsequent horizontal branch stars with the interactions with the environment stars taken into account.  

\acknowledgments

This paper is based on one of the author$^{\prime}$ s(S. Y.) dissertation submitted to Hokkaido University, in partial
fulfillment of the requirement for the doctorate.
  This work has been partially supported by Grant-in-Aid for Scientific Research (15204010, 16540213, 18104003), from Japan Society of the Promotion of Science. 

\appendix

\section{The Polytrope Model of Red Giant Structure }

\citet{fujimoto92} show that the structure of red giant can be modeled by a combination of two polytropes with the cool and hot components corresponding to the core and envelope, respectively;  
   In particular, we may replace the cool component as a sphere of uniform density (i.e., a polytrope of index $N = \infty$) since we are interested only in the envelope structure, and in this case, the structure of hot components ensues from the following equations; 
\begin{eqnarray}
\frac{dM_r}{dr} &=& 4 \pi r^2 \rho, \\
\frac{1}{\rho} \frac{dP}{dr} &=& - \frac{G M_r }{r^2} + g , 
\end{eqnarray}
with the contribution of core gravity, $g$, in eq.~(\ref{eq:red-gravity}) taken into account. 
   With a given envelope mass, $M_{\rm env}$, and the surface radius, $R$, we may introduce the dimensionless variables as 
\begin{eqnarray}
M_r = M_{\rm env} \varphi, \quad r= R \xi, \quad \rho = \rho_0 \theta ^N, \quad P = P_0 \theta^{N + 1}, 
\end{eqnarray} 
   where $\rho_0$ and $P_0$ are the density and pressure coefficients, related to the envelope mass and radius as; 
\begin{eqnarray}
\rho_0 =  M_{\rm env} / 4 \pi R^3, \quad 
P_0 =  G M_{\rm env}^2 / 4 \pi ( N + 1) R^4.  
\end{eqnarray}
   By using these variables, we may rewrite the equations in the non-dimensional form, corresponding to the Lane-Emden equation, as
\begin{eqnarray}
\frac{1}{\xi^2} \frac{d}{d\xi}(\xi^2\frac{d \theta}{{d} \xi}) & = & - \theta ^N - {3}  \frac{{{\varphi_{\rm core}}}}{\xi_{\rm core}^3} \quad \hbox{ for } \xi < \xi_{\rm core} \nonumber \\ 
\frac{1}{\xi^2} \frac{d}{d\xi}(\xi^2\frac{d \theta}{{d} \xi}) & = & - \theta ^N \phantom{- \frac{\phi_{\rm core} \xi}{\xi_{\rm core}^3}} \quad\quad \hbox{for } \xi \ge \xi_{\rm core}, 
\label{eq:dimless-struct}
\end{eqnarray} 
     where $\xi_{\rm core} = R_{\rm core}/ R$ and $\varphi_{\rm core} = M_{\rm core}/ M_{\rm env}$. 
   In the above equations, we take into account the hot component in the core, but its contribution to the envelope mass is negligible because of small core radius.  

We may obtain the envelope structure of red giants by solving eqs.~(\ref{eq:dimless-struct}) for a given set of core radius and mass, ($\xi_{\rm core}, \varphi_{\rm core})$ with the boundary conditions,
\begin{equation}
   d \theta / d \xi =0, \varphi=0  \hbox{ at } \xi = 0:  \hbox{ and } \theta = 0, \varphi = 1 \hbox{ at } \xi = 1.   
\end{equation}
   The several solutions are shown in Fig.~\ref{fig:redgiant}, and we see that the structure, exterior to the core, resemble each other as long as the core radius is sufficiently smaller than the stellar radius ($\xi_{\rm core} \ll 1)$ and unless the envelope mass is much smaller or much larger than the core mass. 
  In particular, we may have relation between the polytropic constant (or entropy) and the surface characteristics of red giant analogous to the single polytrope; 
\begin{equation}
K = P_0/\rho_0^{1+1/N} = \frac{(4 \pi)^{1/N}G}{N} M_{\rm env}^{1-1/N} R^{3/N-1}.
\label{eq:redgiantK} 
\end{equation}

\section{Model Dependences of Transfer of Energy and Angular Momentum due to Linear Dynamical Tide }

The linear perturbation theory of dynamical tides has been developed to estimate the deposition of energy and angular momentum from the orbital motion to the stellar oscillations.  
   \citet{press77} derive an analytic formula for the energy deposition, and later, \citet{lai97} extends it with the stellar rotation taken into account to give the general formulae for both the energy deposition and angular momentum transfer, which are applicable for $\eta \gg 1$.  
   We are here concerned with the red giants in negligible rotation initially. 
   Furthermore, we may well retain only a leading term of $l = 2$, $f$-modes, which dominate the dynamical tides \citep[see][]{lee86}.  

\citet{press77} give the energy loss, $\Delta E$, of the orbital motion into the adiabatic, non-radial oscillations for a star of mass, $M_1$, and radius, $R_1$, during an encounter of periastron distance $r_{\rm p}$ with a point object of mass $M_2$, in the following formua; 
\begin{eqnarray}
\Delta E = \frac{G M_1^2}{R_1} (\frac{M_2}{M_1})^2 \sum_{l=2,3..}(\frac{R_1}{r_{\rm p}})^{2l + 2} T_l (\eta), 
\end{eqnarray} 
   where the dimensionless transfer function, $T_l$, is defined by 
\begin{eqnarray}
T_l(\eta) = 2 \pi^2 \sum_{n}\mid Q_{\rm nl} \mid^2 \sum^l_{m=-l}\mid K_{\rm nlm} \mid^2
\end{eqnarray}
   with the overlap integral, $Q_{\rm nl}$, given by 
\begin{equation}
Q_{\rm nl} = \int^{R_1}_0 r^2 dr \rho l (r / R_1)^{l-1}[{\xi^R}_{\rm nl}  (l){\xi^S}_{\rm nl}], 
\label{eq:q-value}
\end {equation}
   and with the integral, $K_{\rm nlm}$, along the trajectory, given by 
\begin{eqnarray}
K_{\rm nlm} &= & \frac{W_{\rm lm}}{2\pi}2^{3/2}\eta I_{\rm lm}(\eta\omega_{nl}), \\
I_{\rm lm}(y) & = & \int^\infty_0 dx(1+x^2)^{-l}\cos[2^{1/2}y(x + x^3/3) + 2m\hspace{2mm} \tan^{-1}x], \\
W_{\rm lm} & = & (-)^{(l+m)/2}[\frac{4\pi}{2l+1}(l-m)!(l+m)!]^{1/2}/[2^l(\frac{l-m}{2})!(\frac{l+m}{2})!].  
\label{eq:wlm}
\end{eqnarray}  
   Here ${\xi^R}_{\rm nl}$ and ${\xi^S}_{\rm nl}$ are the radial and poloidal normal mode components of the Lagrangian displacements from the unperturbed spherically symmetric state in units of $M_1^{-1/2}$, respectively, and the symbol $(-)^k$ in (\ref{eq:wlm}) is to be interpreted as $(-1)^k$ when $k$ is an integer, while zero when $k$ is not an integer.
  Since $T_l(\eta)$ includes the stellar mode frequency, $\omega_{nl}$, and the overlap integrals, $Q_{\rm nl}$, both determined from the normal mode structures, and hence, $\Delta \tilde{E}$ is dependent on the mode oscillation structure of the star. 

For the leading term of $l=2$, $f$-mode, we have the following form by separating the dependences on the mass and radius of stars as, 
\begin{eqnarray}
& & {\Delta E_2} =  { ({GM_1^2}/{R_1}) ({M_2 / M_1 + M_2})^2  } \eta ^{-4} T_2 (\eta: Q_{02}, \omega_{{02}}), \label{eq:lnfit-energy}\\ 
& & T_2 (\eta: Q_{02}, \omega_{{02}})  =  \frac{4\pi}{5} \mid Q_{02} \mid^2  [I_{20} (\eta \omega_{{02}})^2 + \frac{3}{2}\{I_{22}(\eta \omega_{{02}})^2 + I_{2-2}(\eta \omega_{{02}})^2 \}]. 
\label{eq:transferfunc1}
\end{eqnarray}
    For the red-giants of $M_1 = M_{\rm RG} = 0.8 \msun$, the frequency, $\omega_{{02}}$, and the overlap integral, $Q_{02}$, of $l=2$, $f$-mode oscillations are obtained by \citet{mcmillan90}, who solve the perturbation equations to find the overlap integral and stellar mode frequency.  
   We may evaluate the values of $\omega_{{02}}$ and $Q_{02}$ by interpolating their results as a function of the radius in the following forms;
\begin{eqnarray}
& & \omega_{02} = -1.56 \hbox{E-04}(R_{\rm RG}/R_\sun)^2 + 0.0112 (R_{\rm RG}/R_\sun) + 1.84 . \label{eq:w2int} \\
& & Q_{02} = 2.29 \hbox{E-05}(R_{\rm RG}/R_\sun)^2 - 0.00352 (R_{\rm RG}/R_\sun) + 0.331. 
\label{eq:q2int} 
\end{eqnarray}

As for the angular momentum, $\Delta L$, transferred from the orbit to the spin of primary star, Lai(1997) derives a general form with the effects of stellar rotation taken into account. 
   In the limit of negligible initial rotation rate, it reduces to 
\begin{equation}
\Delta {L} = (GM_1^3 R_1)^{1/2} (\frac{M_2}{M_1})^2 \sum_{l=2,3..}(\frac{R_1}{r_{\rm p}})^{2l + 2} S_l(\eta),
\end{equation}
   where the transfer function, $S_l$, is defined as 
\begin{equation}
S_l(\eta) = -2 \pi^2 \sum_{n} \mid Q_{\rm nl} \mid^2 \omega_{nl}^{-1} \sum^l_{m =-l}m \mid K_{\rm nlm} \mid^2
\end{equation} 
   with the overlap integral, $Q_{\rm nl}$, and the trajectory integral, $K_{\rm nlm}$, defined above.  

Then for the $ l=2$, f-mode, we have
\begin{eqnarray}
& & { \Delta {L}_2} = {\sqrt{G M_1^3 R_1} ({M_2 \over M_1 + M_2})^2} \eta^{-4} S_2 (\eta: Q_{02}, \omega_{{02}}),  \label{eq:lnfit-am} \\
\noalign{where} \\
& & S_2 (\eta: Q_{02}, \omega_{{02}}) =  - (\frac{12 \pi}{5}) \mid Q_{02} \mid ^2 \omega_{{02}}^{-1} \{I_{22}(\eta \omega_{{02}})^2 - I_{2-2}(\eta \omega_{{02}})^2\}].  
\label{eq:transferfunc2}
\end{eqnarray}

\begin{deluxetable}{cccccccccccc}
\tablecaption{Parameters and Characteristic Results of Encounter Simulations. 
}
\tabletypesize{\small}
\tablecolumns{12}
\tablewidth{0pt}
\tablehead {
\colhead{Model} & $\alpha_{\rm SPH}$ & ${M_{\rm MS}}$  &  ${r_{\rm p}}$  & $\eta$ &  ${\Delta E (t_E)}$ &  ${\Delta L_{\rm RG} (t_E)}$ &  $M_{\rm acc} (t_E)$ &  $M_{\rm loss} (t_E)$ &  $T_{\rm orb}$  &  $e$  &  $a$  \\
\colhead{} &  &  $({M_\sun}) $    &  $({R_{\rm RG}}) $ & &  $({GM_{\rm RG}^2 / R_{\rm RG}})$  &   $(M_{\rm RG} R_{\rm RG}^2 \Omega_{\rm K})$  &  $({M_\sun})$  &  $({M_\sun})$  & (yr)  &  &  (au) }
\startdata
&\multicolumn{9}{c} {Red Giant Model of $R_{\rm RG}=20 R_\odot$} & \\
a8rg1 &1.0 &  0.8 & 1.75  & 1.64 & 1.86E-02 &2.6E-02 & 6.70E-03 & 5.09e-04 &   4.19 & 0.95 & 3.04  \\
b8rg1 &1.0 &  0.8  & 2.00  & 2.00  & 8.19E-03&9.2E-03 & 2.91E-03 & 1.92e-05&  22.99 & 0.98  & 9.4  \\     
c8rg1 &1.0 &  0.8 & 2.25  & 2.40 & 3.09E-03 &3.8E-03  &5.47E-04 & 0.0 &  fly-by & - & - \\
d8rg1 &1.0 &  0.8 & 2.50  & 2.80  &  9.48E-04&1.4E-03  &0.0 & 0.0 &  fly-by & - & -  \\ 
e8rg1 &0.1 &  0.8 & 1.75  & 1.64 &  1.88E-02 &2.6E-02 & 6.93E-03 & 5.47e-04 & 4.09   & 0.95 &2.99  \\
f8rg1 &0.1 &  0.8 & 2.00  & 2.00 & 8.32E-03 &9.2E-03 & 2.97E-03 &4.80e-05 & 22.01 & 0.98  & 9.17   \\     
g8rg1 &0.1 & 0.8 & 2.25  & 2.40 &3.19E-03& 3.9E-03&5.37E-04 & 9.60e-06& fly-by & - & -  \\
h8rg1 &0.1 & 0.8 & 2.50  & 2.80 &  1.00E-03 &1.4E-03& 0.0 & 0.0 & fly-by & - & -  \\ 
a6rg1 &1.0 & 0.6 & 1.50  &  1.40& 2.35E-02  &3.3E-02 & 5.81E-03 & 1.14e-03& 1.86 &  0.92  &1.69   \\ 
b6rg1 &1.0 & 0.6 & 1.75  & 1.75 &  1.12E-02 & 1.4E-02 &2.95E-03 & 1.44e-04 &  7.25 & 0.96 & 4.18 \\ 
c6rg1 &1.0 & 0.6 & 2.00  & 2.14 & 4.33E-03 &5.0E-03& 8.06E-04 & 9.60e-06& 91.9 & 0.99 & 22.8 \\
d6rg1 &1.0 & 0.6 & 2.25  & 2.55 & 1.41E-03 &1.9E-03 & 9.60E-06 & 0.0 &  fly-by & -  & -   \\ 
e6rg1 &0.1 & 0.6  & 1.50  & 1.40 &2.37E-02 &3.3E-02 & 5.97E-03 & 0.0 &  fly-by & -   & -  \\
&\multicolumn{9}{c} {Red Giant Model of $R_{\rm RG}=85 R_\odot$} & \\
a8rg2 &1.0 & 0.8& 1.41  & 1.18 &2.49E-02 &3.3E-02 & 6.66E-03 & 2.22E-03 &60.3  & 0.97 &18.0      \\
b8rg2 &1.0 &  0.8 & 1.80  & 1.70 & 9.29E-03 &1.1E-02& 3.20E-03 & 2.50E-04&  fly-by & - & -  \\
c8rg2 &1.0 &  0.8 & 1.88  & 1.82 & 7.18E-03 &9.3E-03& 2.51E-03 & 1.82E-04 &  fly-by & - & -  \\
d8rg2 &1.0 &  0.8 & 2.00  & 2.00 & 4.71E-03 &5.2E-03  & 1.57E-03 & 5.02E-06 & fly-by  & - & -  \\
$e8rg2_{}$ &1.0 &  0.8 & 2.12  & 2.18  &  3.25E-03 &3.5E-03 &  8.83E-04 &  1.28E-05 & fly-by & - & -   \\  
f8rg2 &1.0 &  0.8 & 2.45  & 2.71 &  8.79E-04 &1.1E-03 & 1.92E-05 & 0.0 & fly-by & - & -  \\ 
g8rg2 &0.1 &  0.8 & 1.41  & 1.18 & 2.49E-02 &3.3E-02 & 6.66E-03 & 2.22E-03  &60.3 & 0.97 & 1.80    \\
h8rg2 &0.1 & 0.8 & 1.80  & 1.70 & 9.18E-03 &1.1E-02& 3.52E-03 & 2.24E-04& fly-by & - & - \\
i8rg2 &0.1 & 0.8 & 2.00  & 2.00 &   4.71E-03&5.5E-03 & 1.56E-03 & 6.4E-06 &fly-by & - & -   \\ 
a6rg2 & 1.0 & 0.6 & 1.41  & 1.27  & 1.68E-02  &2.3E-02 &3.56E-03 & 1.13E-03 &142.2  & 0.98 &3.04   \\ 
b6rg2 & 1.0 & 0.6 & 1.65  &1.60 &  8.84E-03 & 1.2E-02 &2.20E-03 & 3.26E-04 &fly-by &- & -\\ 
c6rg2 &1.0 &  0.6 & 1.88 & 1.95 & 4.11E-03&4.8E-03   &4.80E-04 & 2.56E-04 &fly-by &- &- \\
\enddata 
\label{tab:model=alpha}
\end{deluxetable}


\begin{table}
\caption[Table3]{Tidal Capture Limits:\\
First and second columns are main sequence mass and red ginat radius, third and forth ones are $\eta$ and $r_p$ at tidal capture limit for parameters of first and second columns estimated from our simulations and last two columns are same as that of third and forth ones but estimated from linear theory. }
\begin{tabular}{cccccc}
\hline\hline
\hspace{-1mm} $M_{\rm MS} (M_\odot)$ \hspace{-1mm} & \hspace{-1mm} $R_{\rm RG}(R_\sun)$ \hspace{-1mm} & \hspace{-1mm} $ \eta_{\rm cap}$ \hspace{-1mm} & \hspace{-1mm} $r_{\rm p_{\rm cap}}(R_{\rm RG}) $ \hspace{-1mm}  & $ {\eta_{\rm cap}^{\rm linear}}$ \hspace{-1mm} & \hspace{-1mm} ${r_{\rm p_{\rm cap}}^{\rm linear}} (R_{\rm RG}) $  \\
\hline
0.6 &20 & {2.3}  & {2.1}   & {1.78}  & {1.77}  \\
0.8 &20 & {2.38} & {2.25}  & {1.82}  & {1.88}  \\
1.4 &20 & {2.47} & {2.56}  & {1.91}  & {2.16} \\
0.6 &85 & 1.5 & {1.58}   & {1.14}  & {1.31}\\
0.8 &85 & {1.58} & {1.71}  & {1.20}  & {1.42}\\
1.4 &85 & {1.70} & {2.00}  & {1.31}  & {1.68} \\
\hline
\end{tabular}
\label{tb:capturelimit}
\end{table}
\clearpage


\clearpage

\clearpage

\clearpage








\end{document}